%% file: ms.tex
\documentclass[lettersize,journal]{IEEEtran}
\usepackage{amsthm,amsmath,amssymb,amsfonts}
\usepackage{algorithmic}
\usepackage{algorithm}

\usepackage{threeparttable} 
\usepackage{pifont}
\usepackage{booktabs,makecell, multirow, tabularx}
\usepackage{hyperref}
\usepackage{enumitem}
\usepackage{tikz}
\usetikzlibrary{arrows,positioning}
\usepackage{subfigure}
\usepackage{listings}
\usepackage{xcolor}
\usepackage{acro}
\input{acronyms.acn}

\newtheorem{theorem}{Theorem}
\newtheorem{assumption}{Assumption}

\newtheorem{definition}[theorem]{Definition}
\IEEEoverridecommandlockouts

\newcommand{\sys}{QAE-BAC}
\newcommand{\poly}{\operatorname{poly}}
\newcommand{\ZJ}[1]{{\color{black}#1}}
\begin{document}

\title{\sys: Achieving \textbf{Q}uantifiable \textbf{A}nonymity and \textbf{E}fficiency in \textbf{B}lockchain-Based \textbf{A}ccess \textbf{C}ontrol with Attribute
}

\author{
Jie Zhang,~\IEEEmembership{Student Member,~IEEE},
Xiaohong Li,~\IEEEmembership{Member,~IEEE},
Mengke Zhang,
Ruitao Feng,\\
Shanshan Xu,
Zhe Hou, and
Guangdong Bai,~\IEEEmembership{Member,~IEEE}
\thanks{This work is supported in part by the National Key Research and Development Program of China under Grant 2023YFB3107103, in part by the National Natural Science Foundation of China under Grant 62262073, 62332005. }
\thanks{ Jie Zhang, Xiaohong Li and Mengke Zhang are with the College of Intelligence and Computing, Tianjin University, Tianjin, China. (email: \{jackzhang, xiaohongli, mengkezhangcs\}@tju.edu.cn).}
\thanks{ Ruitao Feng is with the Faculty of Science and Engineering, Southern Cross University, Australia (e-mail: ruitao.feng@scu.edu.au).}
\thanks{ Shanshan Xu is with the School of Geographic Sciences, East China Normal University, Shanghai, China. (email: s.xu.ecnu@gmail.com). }
\thanks{ Zhe Hou is with the School of Information and Communication Technology, Griffith University, Nathan, Australia. (email: z.hou@griffith.edu.au).}
\thanks{ Guangdong Bai is with the \ZJ{Department of Computer Science, City University of Hong Kong, Hong Kong, China. (e-mail: baiguangdong@gmail.com)}.}
\thanks{ Jie Zhang and Mengke Zhang contributed equally to this work.}
\thanks{  Ruitao Feng and Guangdong Bai are the corresponding authors.}
}

\markboth{Journal of \LaTeX\ Class Files,~Vol.~14, No.~8, August~2021}%
{Zhang \MakeLowercase{\textit{et al.}}: A Sample Article Using IEEEtran.cls for IEEE Journals}

\IEEEpubidadjcol
\maketitle

\begin{abstract}
\ac{BC-ABAC} offers a decentralized paradigm for secure data governance but faces two inherent challenges: the transparency of blockchain ledgers threatens user privacy by enabling re-identification attacks through attribute analysis, while the computational complexity of policy matching clashes with blockchain's performance constraints. Existing solutions, such as those employing \ac{ZKP}, often incur high overhead and lack measurable anonymity guarantees, while efficiency optimizations frequently ignore privacy implications. To address these dual challenges, this paper proposes \sys{} (\textbf{Q}uantifiable \textbf{A}nonymity and \textbf{E}fficiency in \textbf{B}lockchain-Based \textbf{A}ccess \textbf{C}ontrol with Attribute). \sys{} introduces a formal $(r, t)$-anonymity model to dynamically quantify the re-identification risk of users based on their access attributes and history. Furthermore, it features an \ac{EWPT} that optimizes policy structure based on real-time anonymity metrics, drastically reducing policy matching complexity. Implemented and evaluated on Hyperledger Fabric, \sys{} demonstrates a superior balance between privacy and performance. Experimental results show that it effectively mitigates re-identification risks and outperforms state-of-the-art baselines, achieving up to an 11x improvement in throughput and an 87\% reduction in latency, proving its practicality for privacy-sensitive decentralized applications.
\end{abstract}


\textbf{
	\textcolor{blue!70!black}{
		This work has been submitted to the IEEE for possible publication. Copyright may be transferred without notice, after which this version may no longer be accessible.
}}

\input{01.Introduction.tex} 
\input{02.Related.tex}

\input{03.Preliminaries.tex}
\input{04.Formulation.tex}
\input{05.MainModel.tex}
\input{06.Security.tex}

\input{07.Performance.tex}
\input{09.Conclusion.tex}

\section*{Acknowledgment}
\ZJ{
	The authors would like to thank the anonymous reviewers for their valuable feedback.
	To support reproducible research and foster further development, the source code implementing the \sys{} framework and detailed documentation will be made publicly available on GitHub upon acceptance of this paper.
}

\bibliographystyle{IEEEtran}
\bibliography{IEEEabrv,myrefs}

\input{Appendix.tex}

\end{document}

%% file: acronyms.acn.tex
\DeclareAcronym{ABAC}{
	short=ABAC,
	long=Attribute-Based Access Control,
}
\DeclareAcronym{BC-ABAC}{
	short=BC-ABAC,
	long=Blockchain-based Attribute-Based Access Control,
}
\DeclareAcronym{ZKP}{
	short=ZKPs,
	long=Zero-Knowledge Proofs,
}
\DeclareAcronym{QAE-BAC}{
	short=QAE-BAC,
	long=Quantifiable Anonymity and Efficiency in Blockchain-Based Access Control with Attribute,
}
\DeclareAcronym{EWPT}{
	short=EWPT,
	long=Entropy-Weighted Path Tree,
}
\DeclareAcronym{IoT}{
	short=IoT,
	long=Internet of Things,
}
\DeclareAcronym{AAC}{
	short=AAC,
	long=Anonymous Access Control,
}
\DeclareAcronym{AS}{
	short=AS,
	long=Subject Attributes,
}
\DeclareAcronym{AO}{
	short=AO,
	long=Object Attributes,
}
\DeclareAcronym{AE}{
	short=AE,
	long=Environment Attributes,
}
\DeclareAcronym{OP}{
	short=OP,
	long=Operation Attributes,
}
\DeclareAcronym{DL}{
	short=DL,
	long=Discrete Logarithm,
}
\DeclareAcronym{PPT}{
	short=PPT,
	long=Probabilistic Polynomial-Time,
}
\DeclareAcronym{EUF-CMA}{
	short=EUF-CMA,
	long=Existentially Unforgeable under Chosen Message Attacks,
}
\DeclareAcronym{AQC}{
	short=AQC,
	long=Anonymity Quantification Contract,
}
\DeclareAcronym{AIC}{
	short=AIC,
	long=Attribute Information Contract,
}
\DeclareAcronym{AAH}{
	short=AAH,
	long=Anonymous Access History,
}
\DeclareAcronym{PEC}{
	short=PEC,
	long=Policy Enforcement Contract,
}
\DeclareAcronym{PDC}{
	short=PDC,
	long=Policy Decision Contract,
}
\DeclareAcronym{PAC}{
	short=PDC,
	long=Policy Administration Contract,
}
\DeclareAcronym{OVAT}{
	short=OVAT,
	long=One-Variable-At-A-Time,
}
\DeclareAcronym{ROM}{
	short=ROM,
	long=Random Oracle Model,
}
\DeclareAcronym{ECDSA}{
	short=ECDSA,
	long=Elliptic Curve Digital Signature Algorith,
}
\DeclareAcronym{RQ}{
	short=RQs,
	long=Research Questions,
}

%% file: 01.Introduction.tex
\section{Introduction}
\label{sec:introduction}

The exponential growth of data generation and exchange in modern digital ecosystems, from cloud computing to interconnected \ac{IoT} devices, has made robust access control a critical cornerstone of cybersecurity \cite{IoTreport, IoTdata}. \ZJ{\ac{ABAC}}~\cite{hu2015attribute} has emerged as the preeminent model for managing security in these dynamic and distributed environments. By evaluating policies based on the attributes of subjects, objects, and the environment, \ac{ABAC} provides superior flexibility and fine-grained control compared to traditional role-based models \cite{hao2022blockchain, du2024blockchain, ABAC-RBAC,Zhang2023CrossDomainAC}. However, the conventional deployment of \ac{ABAC} often relies on a centralized authority, introducing critical vulnerabilities like single points of failure and ambiguous data sovereignty \cite{Xia2017BBDSBD,Panda2023ContextualAA,jiang2023sanidea}.

Blockchain technology, with its core tenets of decentralization, immutability, and transparency, presents a compelling solution to this mismatch \cite{hao2022blockchain,bitcoin, Merlec2024SCCAACAS}. By executing \ac{ABAC} policies through smart contracts, researchers have built decentralized access control systems that \ZJ{reduce reliance on a single trusted third party by shifting trust to a decentralized protocol and its immutable code} \cite{islam2019permissioned, Zhang2020AttributeBasedAC, Liu2020FabriciotAB, Wang2022DynamicAC,Duan2023TRCTAT, zhang2025swiftguard}. This fusion, however, intensifies two fundamental and deeply intertwined challenges that threaten the viability of \ZJ{\ac{BC-ABAC}} in practice.

First, the \textbf{privacy-transparency paradox} becomes severe. 
Although the public records of blockchain provide transparency for auditability, this transparency also exposes sensitive attributes required for \ac{ABAC} policy evaluation~\cite{qin2024share,koo2024access}.
The very attributes required for fine-grained policy evaluation (e.g., `role', `clearance', `affiliation') are often sensitive. When recorded on an immutable ledger, they form a rich, permanent dataset for adversaries. Through sophisticated linkage attacks and frequency analysis, malicious actors can de-anonymize users, trace their behavior across transactions, and infer sensitive information \cite{Xia2021EnablingRH, Zhang2024AnonymityIA}. This risk is particularly acute in systems with many fine-grained attributes, where certain combinations can act as quasi-identifiers, uniquely pinpointing individuals within a small user pool \cite{Lanus2023GuaranteeingAI}.

Second, the \textbf{performance-complexity gap} is widened. \ac{ABAC} inherently suffers from the ``attribute explosion" problem~\cite{fedrecheski2021smartabac}, where the growing number and complexity of attributes and policies make the request-to-policy matching process computationally expensive, leading to increased authorization latency \cite{Zhang2022ASA, Lanus2020AlgorithmsFC}. Blockchain platforms, often characterized by lower transaction throughput and higher consensus latency compared to centralized systems, act as a performance bottleneck, dramatically amplifying this inherent complexity \cite{Ullah2023ASO, Tong2022ABD}. Consequently, the combined system struggles to meet the low-latency, high-throughput demands of large-scale, real-world applications \cite{10399957}.

\textbf{Limitations of Existing Work \& Our Motivation:~}Existing research has made considerable strides but often addresses these challenges in isolation, leading to a fragmented landscape. On one hand, \textbf{privacy-focused approaches} frequently employ advanced cryptographic techniques like \ac{ZKP} \cite{zhang2025swiftguard,Wu2022ABB, Hu2024TowardsAA} or anonymization methods \cite{Lanus2023GuaranteeingAI, Yuen2015kTimesAA,fang2025accountable}. While these approaches can hide attribute values or identities, they often introduce substantial computational overhead, lack a mechanism for quantitatively measuring the achieved level of anonymity and may not adequately protect against privacy leaks from dynamic access patterns \cite{Zhang2024AnonymityIA}. On the other hand, \textbf{efficiency-focused approaches} optimize policy retrieval and matching \cite{fedrecheski2021smartabac, Karimi2020AnAA, Geng2023AnAC, BAI2021101957} but are fundamentally \textit{privacy-agnostic}; their design does not consider whether optimizing for speed might inadvertently simplify an attacker's task of re-identifying users, potentially exacerbating the privacy risks they ignore. This clear dichotomy highlights a critical research gap: \textbf{The absence of a \textit{holistic, co-designed framework} where \textit{continuous, quantifiable anonymity assessment actively guides and informs performance optimization}. Without this synergy, systems are forced to choose between privacy and performance, or suffer the penalties of both.} Bridging this gap is the primary motivation for \ZJ{this} work.

\textbf{Our Approach and Novelty:~}This paper proposes \sys{} (\textbf{Q}uantifiable \textbf{A}nonymity and \textbf{E}fficiency in \textbf{B}lockchain-Based \textbf{A}ccess \textbf{C}ontrol with Attribute), a novel framework that breaks the prevailing privacy-efficiency trade-off through deep integration. \ZJ{The core novelty of \sys{} lies in its \textbf{closed-loop feedback system}, which actively uses real-time privacy metrics to govern performance optimization.}
%
This tight coupling ensures that the system is not just fast, but \textbf{responsibly fast}; it is not just private, but \textbf{efficiently private}. \ZJ{The proposed framework} provides a foundational shift towards building scalable, efficient, and truly privacy-preserving decentralized data governance systems.

\textbf{Contributions:~}The main contributions of this work are four-fold:
\begin{itemize}[leftmargin=7.5pt]
	\item \ZJ{The novel \sys{} framework is proposed} to cohesively integrate continuous anonymity quantification with privacy-aware policy optimization for \ac{BC-ABAC}. \ZJ{This deep integration establishes a feedback loop where live anonymity scores directly guide the optimization process,} effectively breaking the prevailing privacy-efficiency trade-off.
	\item \ZJ{The privacy threat in \ac{BC-ABAC} is formalized} by defining a dynamic ``credential subject space" and introducing an $(r, t)$-anonymity model. \ZJ{This model provides a quantitative, real-time metric for assessing re-identification risk~\cite{Xia2021EnablingRH},} addressing a critical gap in existing privacy solutions that offer protection but no measure of its strength.
	\item \ZJ{The \ac{EWPT} structure and a corresponding fast authorization algorithm are designed.} \ZJ{The innovation of \ac{EWPT} is that its weights and structure are derived from real-time anonymity metrics and access patterns, achieving a fundamental reduction in policy matching time complexity ($O(\log n)$) while ensuring that optimization does not create new privacy vulnerabilities.}
	\item \ZJ{A prototype of \sys{} is implemented on Hyperledger Fabric and extensive experiments are conducted} using real-world attribute datasets. Results demonstrate that \sys{} effectively maintains high subject anonymity under various conditions and outperforms state-of-the-art baselines, achieving up to an 11x improvement in throughput and an 87\% reduction in latency.
\end{itemize}

The remainder of this paper is organized as follows. Section \ref{sec:related_work} reviews related literature. Section \ref{sec:preliminaries} presents formal definitions. The overall architecture of \sys{} is described in Section \ref{sec:system_overview}. Section \ref{sec:detailed_design} elaborates on the detailed design of the core modules. Section \ref{sec:security_analysis} provides a security analysis. Performance evaluation is discussed in Section \ref{sec:evaluation}. Finally, Section \ref{sec:conclusion} concludes the paper.

%% file: 02.Related.tex
\section{Related Work}
\label{sec:related_work}
\ZJ{This research} is situated at the intersection of decentralized access control, privacy-enhancing technologies, and efficient policy management. \ZJ{The evolution of relevant research in areas is reviewed}, highlighting the technological advancements and, crucially, the persistent limitations that motivate this work.

\begin{table}[t]
	\centering
	\caption{Comparison of \sys{} with related works.}
	\label{tab:comparison}
	\resizebox{0.5\textwidth}{!}{
		\begin{tabular}{lcccccc}
			\toprule
			\textbf{Work} & \textbf{\makecell[c]{Decentra- \\lized}} & \textbf{\makecell[c]{Fine- \\Grained}} & \textbf{\makecell[c]{Privacy \\ Protection}} & \textbf{\makecell[c]{Quantifiable \\ Anonymity}} & \textbf{\makecell[c]{Efficient \\Authorization}} \\
			\midrule
			\cite{Zhang2020AttributeBasedAC} & \ding{51} & \ding{51} & \ding{55} & \ding{55} & \ding{55} \\
			\cite{Liu2020FabriciotAB} & \ding{51} & \ding{51} & \ding{55} & \ding{55} & \ding{55} \\
			\cite{Tong2022ABD} & \ding{51} & \ding{55} & \ding{55} & \ding{55} & \ding{51} \\
			\cite{Zhang2023CrossDomainAC} & \ding{51} & \ding{51} & \ding{55} & \ding{55} & \ding{51} \\
			\cite{Wu2022ABB} & \ding{51} & \ding{51} & \ding{51} (ZKP) & \ding{55} & \ding{55}\\
			\cite{Hu2024TowardsAA} & \ding{51} & \ding{51} & \ding{51} (ZKP) & \ding{55} & \ding{55} \\
			\cite{fang2025accountable} & \ding{51} & \ding{51} & \ding{51} (AAC) & \ding{55} & \ding{55} \\
			\cite{Lanus2023GuaranteeingAI} & \ding{55} & \ding{51} & \ding{51} (AAC) & - & \ding{55}\\
			\cite{Zhang2024AnonymityIA} & \ding{55} & \ding{51} & \ding{51}(AAC) & \ding{51} & \ding{55} \\
			\cite{Geng2023AnAC} & \ding{55} & \ding{51} & \ding{55} & \ding{55} & \ding{51}\\
			\midrule
			\textbf{\sys{}} & \ding{52} & \ding{52} & \ding{52}(AAC) & \ding{52} & \ding{52}\\ \midrule
			\multicolumn{7}{l}{ \ding{51}: supported, \ding{55}: not supported, -: unknown, \ac{AAC}} \\
			\bottomrule
		\end{tabular}
	}
\end{table}

\begin{table}[t]
	\centering
	\caption{Qualitative and Quantitative Comparison of Privacy-Efficiency Trade-off}
	\label{tab:comparison_quantitative}
	\resizebox{0.5\textwidth}{!}{
		\begin{tabular}{lcccl}
			\toprule
			\textbf{Approach} & \textbf{Privacy} & \textbf{Overhead} & \textbf{Complexity} & \textbf{Limitation} \\
			\midrule
			\textbf{\ac{BC-ABAC}~\cite{Liu2020FabriciotAB}} & Low & Low & $O(n)$ & \makecell[l]{High re-identification\\ risk, attributes exposed.} \\
			\textbf{ZKP-based~\cite{Hu2024TowardsAA}} & \makecell[c]{Very High \\ (Value Hiding)} & Very High & \makecell[c]{$O(n)$ + \\(ZKP cost)} & \makecell[l]{High latency, no \\ quantifiable measure.} \\
			\textbf{\ac{AAC}~\cite{Zhang2024AnonymityIA}} & \makecell[c]{High \\(Issuance)} & Medium & $O(n)$ & \makecell[l]{Anonymity degrades\\ with history.} \\
			\textbf{\makecell[l]{Policy Mana.~\cite{Geng2023AnAC} \\ Optimization}} & Low & Low & $\sim O(\log n)$ & \makecell[l]{Privacy-agnostic, may \\reduce anonymity.} \\
			\midrule
			\textbf{\sys{}} & \textbf{High (Quant.)} & \textbf{Medium} & \textbf{$O(\log n)$} & \textbf{\makecell[l]{Balances privacy \\ and performance.}} \\
			\bottomrule
		\end{tabular}
	}
\end{table}

\subsection{The Evolution of Blockchain-Based Access Control}
The exploration of blockchain for access control has been driven by the need to overcome the limitations of centralized architectures. Initial research focused on \textit{demonstrating feasibility}. Works like \cite{Zhang2020AttributeBasedAC} and \cite{Liu2020FabriciotAB} implemented core \ac{ABAC} functionalities—such as policy management and attribute verification—using smart contracts on platforms like Ethereum and Hyperledger Fabric. These pioneering studies proved the concept of decentralized, trustworthy authorization but also revealed significant challenges, primarily concerning on-chain storage costs and the privacy risks inherent in storing attributes on a transparent ledger.

Subsequent efforts focused on \textit{architectural optimizations} to improve scalability. A common strategy involved hybrid storage models \cite{Xia2017BBDSBD, Yang2024AnOE, Zhang2022DistributedSS}, where only hashes or references are stored on-chain while sensitive data is kept off-chain (e.g., in cloud storage). While this alleviated storage pressure, a fundamental privacy issue remained largely unaddressed: the attributes used in on-chain policy evaluation logic itself can leak sensitive information. Another line of research focused on \textit{structural scalability} for cross-domain scenarios. Proposals included sharding architectures \cite{Tong2022ABD} and master-slave chain models \cite{Zhang2023CrossDomainAC,zhang2025swiftguard}, which improved throughput and interoperability between domains. However, these works primarily prioritized performance and architectural design, often overlooking the nuanced privacy implications arising from attribute exposure during cross-domain interactions.

A persistent challenge that spans these architectural evolutions is the ``attribute explosion" problem \cite{Biswas2017AttributeTF}. As systems scale, the proliferation of attributes and policies leads to increased computational complexity during policy matching, creating a performance bottleneck that blockchain's inherent latency further aggravates~\cite{Ullah2023ASO,dong2023blockchain}. While the aforementioned works made systems more scalable, they did not fundamentally solve the core performance-complexity issue within the \ac{ABAC} engine itself. \sys{} addresses this by proposing a novel policy structuring mechanism called \ZJ{\ac{EWPT}} that tackles the root cause of matching inefficiency.


\subsection{Privacy Enhancements in Access Control}
The transparency of blockchain has spurred significant research into privacy-preserving techniques for access control. These efforts can be broadly categorized into two strands.

The first strand focuses on \textit{cryptographic solutions}, most notably \ZJ{\ac{ZKP}} \cite{zhang2025swiftguard, Wu2022ABB, Hu2024TowardsAA}. \ac{ZKP} allow a user to prove the validity of a claim (e.g., that their attributes satisfy a policy) without revealing any underlying information~\cite{Lin2023AnAC}. This provides a strong guarantee for hiding attribute values. However, this power comes at a high computational cost, often adding substantial overhead and latency to the authorization process. Moreover, \ac{ZKP} protect the ``data" but not the ``metadata"; the structure of the policies and the mere fact of an access request can still leak information \cite{Santis2024AnIA}. Most critically, ZKP-based schemes lack a mechanism to \textit{quantify the level of anonymity} they provide, offering protection without measurement.

The second strand employs \textit{anonymization techniques}, often through \ZJ{\ac{AAC}} and credential systems \cite{Yuen2015kTimesAA,fang2025accountable}. These approaches decouple a user's identity from their access rights by using anonymous credentials. Advanced schemes incorporate k-anonymity \cite{Lanus2023GuaranteeingAI} or similar principles to ensure that a user cannot be distinguished within a group of a certain size. However, a common limitation of these approaches is their often ``static nature". They excel at protecting privacy during credential issuance but may not adequately account for the gradual erosion of anonymity that can occur over time through the correlation of ``dynamic access patterns" \cite{Zhang2024AnonymityIA}. Some recent work \cite{Zhang2024AnonymityIA} has begun to consider the impact of ongoing access requests on anonymity but may not fully capture the distribution of attributes in the subject space.

The key limitation across both strands is the \textbf{lack of a dynamic, quantifiable metric for anonymity}. Protection is offered, but its degree and evolution over time are not measured. \sys{} addresses this by introducing a continuous $(r, t)$-anonymity assessment, providing a measurable foundation for privacy that can adapt to the evolving state of the system.


\subsection{Efficient Policy Management}
Managing the complexity of \ac{ABAC} policies is a long-standing research problem. To address the performance degradation caused by ``attribute explosion"\cite{Biswas2017AttributeTF}, researchers have investigated automated policy learning from access logs \cite{Karimi2020AnAA}, conflict resolution algorithms \cite{Geng2023AnAC}, and optimized data structures for policy retrieval \cite{BAI2021101957}. While these methods can improve efficiency, they are typically designed for and evaluated in centralized environments. More importantly, they are largely ``privacy-agnostic". The optimization goals are solely based on performance metrics like speed or storage, with no consideration for whether the resulting policy organization might inadvertently make re-identification easier for an adversary (e.g., by creating efficient lookup paths for unique attribute combinations).

This creates a significant risk: a system could be optimized for performance at the direct expense of privacy. \sys{} innovates by making policy optimization ``privacy-aware". The \ZJ{\ac{EWPT}} is structured based on entropy weights derived from the anonymity assessment module, ensuring that gains in efficiency are achieved in a way that is harmonious with, and even enhances, the privacy goals of the system.


\subsection{Comparison with State-of-the-Art}
Table \ref{tab:comparison} summarizes the comparison between \sys{} and related works across key dimensions. It highlights that \sys{} uniquely combines quantifiable anonymity assessment with privacy-aware policy optimization for efficient authorization in \ZJ{\ac{BC-ABAC}}, addressing gaps left by prior works. Furthermore, Table \ref{tab:comparison_quantitative} provides a qualitative and quantitative comparison, underscoring \sys{}'s advantage in achieving a superior privacy-efficiency trade-off, a balance that existing solutions fail to strike.

%% file: 03.Preliminaries.tex
\section{Preliminaries}
\label{sec:preliminaries}

This section formalizes the core concepts underpinning the \sys{} framework, establishing the foundational terminology for this subsequent discussion on anonymity quantification and access control mechanisms. 

\subsection{Formal Model of Attribute-Based Access Control}
\label{subsec:abac_model}

This subsection introduces the basic elements of attribute-based access control model~\cite{hu2015attribute}, including entities, credentials, requests, and policies, which collectively define the structure and behavior of the access control system.

\begin{definition}[Attribute Space]
	The attribute space $A$ encompasses the entire set of attributes utilized in access control decisions. It is composed of subject attributes ($AS$), object attributes ($AO$), environment attributes ($AE$), and operation attributes ($OP$), denoted as $A = AS \cup AO \cup AE \cup OP = \{a_1, a_2, \cdots, a_k\}$. Each attribute $a \in A$ is defined as a tuple $a = (t, w, V)$, where $t$ denotes the attribute type, $w$ represents its weight (signifying its importance in authorization), and $V$ is the domain of its possible values. When an entity is assigned an attribute $a$, its value is denoted as $v_i \in V$.
\end{definition}

\begin{definition}[Subject]
	A subject $s$ is an entity capable of initiating access requests. It is represented by the values of its assigned subject attributes, i.e., $s = (v_1, v_2, \cdots, v_k)$, where $\forall v_i, \exists a_i \in AS$ such that $v_i \in V_i$. The access permissions of a subject are determined by whether its attribute assignment satisfies the governing access policies.
\end{definition}

\begin{definition}[Object]
	An object $o$ is a resource that is the target of an access request. It is similarly represented by a tuple of its attribute values, $o = (v_1, v_2, \cdots, v_k)$, where $\forall v_i, \exists a_i \in AO$ such that $v_i \in V_i$. Access to an object is granted based on the requesting subject's attributes and the object's own attributes.
\end{definition}

\begin{definition}[Attribute Credential]
	\label{def:attribute_credential}
	An attribute credential $c$ is a minimal subset of a subject's attributes presented during an access request for the purposes of authentication and authorization. Formally, for a subject $s$, a credential is defined as $c \subseteq s$ and $|c| \leq |s|$. Using a minimal credential mitigates privacy risks by limiting the exposure of unnecessary attribute information.
\end{definition}

\begin{definition}[Access Request]
	\label{def:access_request}
	An access request $req$ is the fundamental unit upon which an authorization decision is made. It is a tuple composed of the subject's attribute credential, the target object, the requested operation, and the environmental context: $req = (c, o, op, env)$.
\end{definition}

\begin{definition}[Access Policy]
	\label{def:access_policy}
	An access policy $P$ is a set of rules governing permissions to resources, defined as $P = \{r_1, r_2, \cdots, r_n\}$. Each rule $r_i$ is a conjunction of attribute constraints, expressed as a set of attribute-value pairs: $r_i = {(a_1:v_1), (a_2:v_2), \cdots, (a_m:v_m)}$, where ${a_1, a_2, \cdots, a_m} \subseteq A$ are the attributes constrained by the rule, and ${v_1, v_2, \cdots, v_m}$ are their required values. A request $req$ is granted if it satisfies all constraints of at least one rule $r_i \in P$.
\end{definition}

\subsection{Anonymity Metrics and Quantification}
\label{subsec:anonymity_metrics}

This subsection formalizes the concepts of identifiability and anonymity quantification, providing the mathematical foundation for measuring privacy protection in this framework. The definitions establish how the model adversarial uncertainty and quantify anonymity levels.

\begin{definition}[Subject Identifier]
	\label{def:sub_identifier}
	An attribute or a combination of attribute values that uniquely identifies a single subject $s$ within the subject space $S$ is termed a subject identifier. \ZJ{Identifiers are classified} into two types:
	\begin{itemize}[leftmargin=7.5pt]
		\item \textbf{Explicit Identifier}: An attribute (e.g., a unique ID) that is directly and unambiguously bound to a subject's identity, forming a bijection with the subject space. Administrators typically use explicit identifiers for subject management.
		\item \textbf{Implicit Identifier}: A set of one or more attribute values whose specific combination points uniquely to a single subject, even though the individual attributes themselves are not exclusive identifiers. A subject $s_i$ can have $k$ implicit identifiers, where $k \in [0, 2^{|s_i|} - 1]$. The presence of an implicit identifier in an access request poses a significant re-identification risk \cite{Zhang2024AnonymityIA}.
	\end{itemize}
\end{definition}

\begin{definition}[Credential Subject Space]
	\label{def:credential_subject_space}
	Given credential $c$, its subject space $\mathcal{SS}_{c}$ contains subjects potentially linked to requests carrying $c$:
	\begin{align}
		\mathcal{S}_1^c &= \{ s \mid c \subseteq s,\ s \in S \}
		\quad \text{(Generators of $c$)} \\
		\mathcal{S}_2^c &= \left\{ s \mid
			\begin{aligned}[t]
				&\exists req \in R, \\
				&c \in req \land c \subseteq s
			\end{aligned}
			\right\}
		\quad \text{(Historical users of $c$)} \\
		\mathcal{SS}_c &= \mathcal{S}_1^c \cup \mathcal{S}_2^c
	\end{align}
	where $S$ is the subject space and $R$ the historical request set.
	Anonymity quantification depends on $\mathcal{SS}_c$'s distribution \cite{Zhang2024AnonymityIA}.
\end{definition}

\begin{definition}[Request Probability Entropy]
	\label{def:request_entropy_formal}
	For a request $req_c$ carrying signed credential $\sigma_s(c)$, let $X$ be the random variable representing an adversary's guess of the subject origin. The request probability entropy $\mathcal{E}{req}(req_c)$ is defined as:
	\begin{equation}
		\mathcal{E}{req}(req_c) = -\sum_{s \in \mathcal{SS}_c} P(X=s) \log_2 P(X=s)
	\end{equation}
	where $P(X=s)$ is the probability that subject $s$ originated the request, typically estimated by the relative frequency of $s$ in $\mathcal{SS}_c$ or using a uniform prior distribution.
\end{definition}
The request entropy $\mathcal{E}{req}(req_c)$ measures the adversary's uncertainty in bits, with higher values indicating stronger anonymity protection.

\begin{definition}[$(r,t)$-Anonymity]
	\label{def:rt_anonymity}
	Given a subject attribute distribution matrix $M$ with $N$ subjects and $k$ attributes, let $t$ be the number of non-empty attributes a subject possesses. The matrix $M$ is said to satisfy $(r,t)$-anonymity if for the set of subjects $a_t = { s_i \mid t \leq |s_i|, s_i \in M }$ (subjects with at least $t$ attributes), the minimum size of the credential subject space $\mathcal{S}_1^{s_i}$ (for any $s_i \in a_t$) is at least $r$:
	\begin{align}
		a_t = { s_i \mid t \leq |s_i|, s_i \in M } \\
		r = \min { |\mathcal{S}_1^{s_i}| \mid s_i \in a_t }
	\end{align}
	This metric, adapted from \cite{Lanus2020AlgorithmsFC}, provides a worst-case guarantee against re-identification for subjects with a given number of attributes. A higher $r$ indicates a stronger anonymity level for the subject space configuration.
\end{definition}

\subsection{Cryptographic Foundation}
\label{subsec:crypto_foundation}

The \sys{} framework is built upon a rigorous cryptographic foundation to ensure tamper-resistant and verifiable security guarantees. During registration, each subject $s$ is assigned a public-private key pair $(pk_s, sk_s)$. The system's security rests on the following standard computational hardness assumption:

\begin{assumption}[\ac{DL} Assumption]\label{ass:dl}
	Let $\mathbb{G}$ be a cyclic group of prime order $p$ with generator $g$. For a uniformly random element $h = g^a \in \mathbb{G}$ where $a \xleftarrow{\$} \mathbb{Z}_p$, no \ac{PPT} algorithm $\mathcal{A}$ can recover the exponent $a$ with probability greater than $\mathsf{negl}(\lambda)$, where $\lambda$ is the security parameter.
\end{assumption}

\ZJ{A digital signature scheme that is \ac{EUF-CMA}~\cite{DBLP:journals/joc/BellareNN09} is employed}, whose security is reducible to the \ac{DL} assumption. This scheme enables subjects to generate unforgeable attestations for their attribute credentials.

%% file: 04.Formulation.tex
\section{System Overview}
\label{sec:system_overview}
This section delineates the overarching architecture and operational workflow of the proposed \sys{} framework. Designed as an enhancement to the traditional \ac{ABAC} model, \sys{} is architected to mitigate identity re-identification risks and alleviate authorization inefficiencies endemic to large-scale, dynamic systems. 

\subsection{Threat Model and Design Goals}
\label{subsec:threat_model}

To rigorously evaluate the security of the \sys{} framework, \ZJ{a comprehensive threat model is first established, core design goals are defined, and formal security definitions are provided. These elements are discussed and analyzed in Section~\ref{sec:detailed_design} and \ref{sec:security_analysis}.}

\subsubsection{Threat Model}
\ZJ{A powerful adversary $\mathcal{A}$ is considered,} who can perform both passive and active attacks~\cite{Dolev-Yao} within the capabilities in a blockchain environment. The capabilities and limitations of $\mathcal{A}$ are as follows:
\begin{itemize}[leftmargin=7.5pt]
	\item \textbf{Capabilities:}
	\begin{enumerate}[leftmargin=7.5pt,label=(\roman*)]
		\item \textbf{Eavesdropping:} $\mathcal{A}$ can observe all network traffic, including access requests, policy decisions, and on-chain transactions, gaining full knowledge of the attribute space $A$ and the structure of access policies $P$.
		\item \textbf{Analysis:} $\mathcal{A}$ can perform long-term, large-scale data analysis and linkage attacks on the publicly available or intercepted data. This includes analyzing the frequency and co-occurrence of attributes to infer sensitive information.
		\item \textbf{Compromise:} $\mathcal{A}$ may compromise a limited number of subjects (users) to learn their full attribute sets and use them to launch more targeted attacks. $\mathcal{A}$ may also collude with other malicious entities.
		\item \textbf{Query:} $\mathcal{A}$ can actively initiate access requests to probe the system, observing the authorization outcomes to learn about policy rules and attribute distributions.
	\end{enumerate}
	\item \textbf{Limitations:}
	\begin{enumerate}[leftmargin=7.5pt,label=(\roman*)]
		\item $\mathcal{A}$ is computationally bounded and cannot break standard cryptographic primitives (e.g., hash functions, digital signatures) used by the underlying blockchain platform (Hyperledger Fabric).
		\item $\mathcal{A}$ does not control the majority of the blockchain network's consensus power and cannot arbitrarily tamper with or revert confirmed transactions.
		\item $\mathcal{A}$ cannot directly compromise the integrity of the smart contracts (chaincode) once they are deployed and running correctly.
	\end{enumerate}
\end{itemize}

The primary security objective is to protect the \textit{identity privacy} of honest subjects. Specifically, $\mathcal{A}$'s goal is to successfully re-identify the subject $s$ behind an access request $req$ or to link multiple requests to the same subject, even when the requests use different attribute credentials.

\subsubsection{Design Goals}\label{designgoals}
The \sys{} framework is designed to achieve the following goals under the aforementioned threat model:
\begin{description}
	\item[\textbf{G1}] \textbf{Subject Anonymity}: The system should ensure that an adversary cannot determine the real-world identity of a subject from its attribute credential $c$ or a series of credentials used in access requests. This requires that for any credential $c$, the credential subject space $\mathcal{SS}_c$ is sufficiently large and non-unique.
	\item[\textbf{G2}] \textbf{Unlinkability}: Given two or more access requests, an adversary should not be able to determine with confidence whether they originated from the same subject, unless this is explicitly revealed by the policy logic itself.
	\item[\textbf{G3}] \textbf{Fine-Grained \& Efficient Access Control}:The system must enforce fine-grained access control policies without compromising performance. Authorization decisions should be both accurate and efficient, even as the number of subjects, attributes, and policies scales.
	\item[\textbf{G4}] \textbf{Resilience to Attribute Correlation Attacks}: The system should be resilient against attacks that leverage the correlation between different attributes or between requests and background knowledge to reduce anonymity.
\end{description}

\subsubsection{Security Definitions}
Based on the threat model and design goals, \ZJ{the key security properties of \sys{} are formalized.}

\begin{definition}[Request Anonymity]\label{def:request_anonymity}
	Let $\Pi$ be the \sys{} framwork. Let $\mathcal{A}$ be a probabilistic polynomial-time (PPT) adversary. Consider the following experiment $\textsf{Exp}_{\mathcal{A},\Pi}^{\textsf{Req-Anon}}(1^\lambda)$:
	\begin{enumerate}[label=(\roman*)]
		\item The system is setup, and $\mathcal{A}$ is given full knowledge of the public parameters and the attribute space $A$.
		\item $\mathcal{A}$ chooses two subjects $s_0, s_1 \in S$ that possess a valid credential $c$ for a target object $o$ and operation $op$.
		\item A bit $b \xleftarrow{\$} \{0,1\}$ is chosen uniformly at random. $\mathcal{A}$ is given a request $req_b$ generated by $s_b$ for $(o, op)$.
		\item $\mathcal{A}$ outputs a guess bit $b'$.
	\end{enumerate}
	The advantage of $\mathcal{A}$ is defined as:
	\[
	\textsf{Adv}_{\mathcal{A},\Pi}^{\textsf{Req-Anon}} = \left| \Pr[b' = b] - \frac{1}{2} \right|
	\]
	Framwork $\Pi$ provides \textit{Request Anonymity} if for all PPT adversaries $\mathcal{A}$, $\textsf{Adv}_{\mathcal{A},\Pi}^{\textsf{Req-Anon}}$ is negligible in the security parameter $\lambda$.
\end{definition}

\begin{definition}[Request Unlinkability]\label{def:request_unlinkability}
	Let $\Pi$ be the \sys{} framwork. Let $\mathcal{A}$ be a PPT adversary. Consider the following experiment $\textsf{Exp}_{\mathcal{A},\Pi}^{\textsf{Unlink}}(1^\lambda)$:
	\begin{enumerate}[label=(\roman*)]
		\item The system is setup.
		\item $\mathcal{A}$ selects a subject $s$ and observes a sequence of $n$ requests $\{req_1, ..., req_n\}$ from $s$.
		\item $\mathcal{A}$ is then presented with two new requests $(req_a, req_b)$, where one is from $s$ and the other is from a randomly chosen different subject $s'$ that can generate a request for the same $(o, op)$. $\mathcal{A}$ must determine which request belongs to $s$.
	\end{enumerate}
	The advantage of $\mathcal{A}$ is defined analogously to Definition \ref{def:request_anonymity}. Framwork $\Pi$ provides \textit{Request Unlinkability} if this advantage is negligible for all \ac{PPT} $\mathcal{A}$.
\end{definition}

\subsection{Core Modules}
\label{subsec:core_modules}

\begin{figure}[h]
	\centering
	\includegraphics[width=0.5\textwidth]{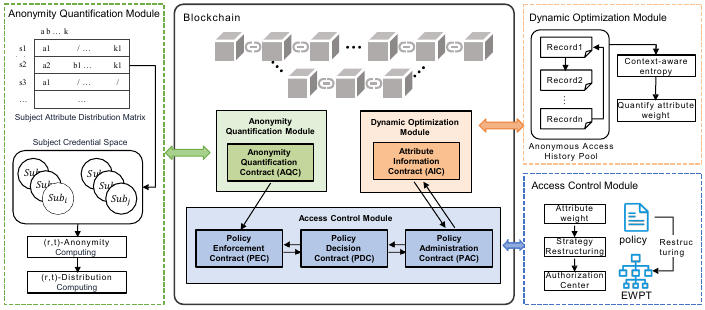}
	\caption{Architecture of the \sys{} framework.} 
	\label{fig:system1}
\end{figure}

The architecture of \sys{}, illustrated in Fig.~\ref{fig:system1}, consists of three core modules deployed on the blockchain. A key enhancement across these modules is the integration of cryptographically signed attribute credentials.

\subsubsection{Anonymity Quantification Module}
This module evaluates request and subject anonymity, forming the framework's privacy foundation through its implementation in the \textbf{\ac{AQC}}. To mitigate re-identification risks from attribute combinations, the \ac{AQC} first requires and verifies a \textit{signed credential} $\sigma_s(c)$ against the purported subject's public key $pk_s$ before any anonymity computation. This critical enhancement ensures credential authenticity and prevents forgery attacks. Only after cryptographic validation does the \ac{AQC}:
\begin{itemize}
	\item Analyze the subject attribute matrix
	\item Construct credential subject spaces $\mathcal{SS}_c$ (Def.~\ref{def:credential_subject_space})
	\item Apply $(r,t)$-anonymity \cite{Zhang2024AnonymityIA,Lanus2020AlgorithmsFC} to quantify anonymity
\end{itemize}
Requests meeting predefined anonymity thresholds are forwarded to policy execution.

\subsubsection{Dynamic Optimization Module}
Replacing conventional Policy Information Points, this module operates via the \textbf{\ac{AIC}} and introduces dynamic attribute weighting through two key mechanisms:
\begin{itemize}
	\item Maintaining an \textit{\ac{AAH} Pool} recording recent requests and authorization outcomes
	\item Computing context-aware entropy by combining \ac{AAH} data with \ac{AQC} anonymity scores
\end{itemize}
The output is an optimized attribute weight list, prioritized by discriminatory power, which enhances policy evaluation efficiency when fed to the access control module.


\subsubsection{Access Control Module}
As the core authorization engine, this module refines traditional \textbf{\ac{PEC}, \ac{PDC}, \ac{PAC}} through three smart contracts. A key security enhancement requires the \ac{PEC} to first verify the signature $\sigma_s(c)$ upon receiving requests $req = (\sigma_s(c), o, op, env)$, ensuring requests originate from legitimate subjects holding corresponding private keys. Only after successful validation does the system:
\begin{itemize}
	\item Parse requests and validate credentials via \ac{PEC}
	\item Dynamically reconstruct the \textit{\ac{EWPT}} via \ac{PAC}
	\item Perform fast authorization through path existence checks
\end{itemize}
The EWPT structure addresses efficiency challenges in high-dimensional spaces by restructuring flat policy rules (Def.~\ref{def:access_policy}) into weight-indexed trees.

\subsection{Operational Workflow}
\label{subsec:workflow}

\ZJ{The end-to-end workflow of \sys{} is composed of four sequential phases that incorporate cryptographic verification.}

\textbf{Phase 1: Anonymity Assessment.} Executed by the \ac{AQC}, this phase establishes the privacy baseline. Steps include: (1) configuring the attribute space $A$; (2) registering subjects/objects and assigning attributes; (3) for an access request with signed credential $\sigma_s(c)$, the \ac{AQC} first verifies the signature. If valid, it computes the request probability entropy and $(r,t)$-anonymity to determine if the request meets anonymity thresholds.

\textbf{Phase 2: Request Authentication.} Handled by the \ac{PEC}, this phase involves two verifications: (1) cryptographic validation of $\sigma_s(c)$; and (2) confirmation that the request passed Phase 1. Requests failing either check are rejected immediately.

\textbf{Phase 3: Dynamic Authorization.} For authenticated requests, the \ac{PDC} checks the \ac{EWPT} for a path matching the request's attribute sequence. Authorization is granted if a valid path exists, ensuring efficient and fine-grained access control.

\textbf{Phase 4: Weight Update.} Operated by the \ac{AIC} on a periodic basis, this phase updates attribute weights using context-aware entropy from the \ac{AAH} Pool and anonymity scores from the \ac{AQC}. The updated weights are pushed to the \ac{PAC} to reorganize the \ac{EWPT}, completing the feedback loop for continuous optimization.

%% file: 05.MainModel.tex
\section{Detailed Design of \sys{}}
\label{sec:detailed_design}

This section presents the comprehensive algorithmic foundation of the \sys{} framework.\ZJ{The eight core algorithms that implement the three functional modules are detailed,} integrating the cryptographic foundation from Section~\ref{subsec:crypto_foundation} with the information-theoretic anonymity metrics \ZJ{that form the core of the privacy preservation approach.}
The algorithms are designed to work in concert to achieve our four primary goals (\textbf{G1-G4}) in Section~\ref{designgoals}.

\subsection{Anonymity Quantification Module}
\label{subsec:anon_quantification_design}

The Anonymity Quantification Module provides formal, measurable guarantees against re-identification attacks. Its design is grounded in information theory, using entropy to quantify the uncertainty an adversary faces when attempting to identify subjects from their attribute credentials. The module comprises four algorithms that collectively ensure the achievement of \textbf{G1} and \textbf{G2} by rigorously measuring and enforcing anonymity levels.

Algorithm~\ref{alg:credential-verification} is the foundational step for all subsequent anonymity calculations. It first verifies the cryptographic signature on the credential to ensure its authenticity and integrity, preventing forgery and spoofing attacks. This step is critical for maintaining the trustworthiness of the system. It then constructs the \textit{credential subject space} $\mathcal{SS}_c$ by combining two sets: subjects who can generate the credential ($\mathcal{S}_1^c$) and subjects who have used it historically ($\mathcal{S}_2^c$). The size of $\mathcal{SS}_c$ directly determines the theoretical upper bound of anonymity for the request. This algorithm directly contributes to \textbf{G1} and \textbf{G2} by ensuring that only valid credentials are processed and by defining the population of possible subjects for a given credential, which is essential for quantifying anonymity.
\begin{algorithm}[t]
	\caption{Credential Verification and Subject Space Construction}
	\label{alg:credential-verification}
	\begin{algorithmic}[1]
		\REQUIRE Signed credential $\sigma_s(c)$, subject space $S$, request history $R$, public key $pk_s$
		\ENSURE Credential subject space $\mathcal{SS}_c$ or $\mathsf{ERROR}$
		\STATE \textbf{Verify Signature:} $\mathsf{result} \gets \mathsf{Verify}(pk_s, c, \sigma_s(c))$
		\IF{$\mathsf{result} = \mathsf{False}$}
		\RETURN $\mathsf{ERROR}$ \COMMENT{Reject forged or tampered credential}
		\ENDIF
		\STATE $\mathcal{S}_1^c \gets \{ s \mid c \subseteq s, s \in S \}$ \COMMENT{Construct capability set}
		\STATE $\mathcal{S}_2^c \gets \{ s \mid \exists req \in R \text{ where } c \in req \wedge s \text{ initiated } req \}$ \COMMENT{Construct usage set}
		\STATE $\mathcal{SS}_c \gets \mathcal{S}_1^c \cup \mathcal{S}_2^c$
		\IF{$|\mathcal{SS}_c| = 0$}
		\RETURN $\mathsf{ERROR}$ \COMMENT{Invalid credential configuration}
		\ENDIF
		\RETURN $\mathcal{SS}_c$
	\end{algorithmic}
\end{algorithm}

Algorithm~\ref{alg:request-anonymity} computes the request probability entropy $\mathcal{E}_{req}(req_c)$ based on the credential subject space $\mathcal{SS}_c$ generated by Algorithm~\ref{alg:credential-verification}. The entropy is calculated using the Shannon entropy formula, which measures the adversary's uncertainty about the subject's identity. A higher entropy value indicates greater anonymity. This algorithm is the core metric for evaluating \textbf{G1} at the request level. It also supports \textbf{G2} by ensuring that multiple requests from the same subject, when using different credentials, yield high entropy values, making linking difficult. The algorithm returns zero if $\mathcal{SS}_c$ has only one subject, indicating a complete loss of anonymity.
\begin{algorithm}[t]
	\caption{Request Anonymity Quantification}
	\label{alg:request-anonymity}
	\begin{algorithmic}[1]
		\REQUIRE Credential subject space $\mathcal{SS}_c$, request $req_c$
		\ENSURE Request anonymity metric $\mathcal{E}_{req}(req_c)$
		\STATE $\mathcal{E}_{req} \gets 0.0$
		\IF{$|\mathcal{SS}_c| = 1$}
		\RETURN $0.0$ \COMMENT{Zero anonymity - implicit identifier}
		\ENDIF
		\FORALL{subject $s_i \in \mathcal{SS}_c$}
		\STATE $p_i \gets \frac{\text{frequency of } s_i \text{ in } \mathcal{SS}_c}{\sum_{s_j \in \mathcal{SS}_c} \text{frequency of } s_j}$ \COMMENT{Probability estimation}
		\STATE $\mathcal{E}_{req} \gets \mathcal{E}_{req} - p_i \cdot \log_2(p_i)$ \COMMENT{Accumulate entropy}
		\ENDFOR
		\RETURN $\mathcal{E}_{req}$
	\end{algorithmic}
\end{algorithm}

Algorithm~\ref{alg:rt-anonymity} evaluates the systemic anonymity of the entire subject population using the $(r,t)$-anonymity model. It iterates over all subjects with at least $t$ attributes and computes the minimum credential subject space size $r$ across these subjects. This algorithm provides a global view of anonymity, ensuring that even subjects with many attributes are protected by a sufficiently large anonymity set. It directly contributes to \textbf{G1} by guaranteeing a baseline level of anonymity for all subjects and to \textbf{G4} by ensuring that attribute combinations do not easily lead to re-identification. The algorithm relies on Algorithm~\ref{alg:credential-verification} to compute the subject space for each subject.
\begin{algorithm}[t]
	\caption{$(r,t)$-Anonymity Assessment}
	\label{alg:rt-anonymity}
	\begin{algorithmic}[1]
		\REQUIRE Subject attribute matrix $M$, attribute threshold $t$
		\ENSURE $(r,t)$-anonymity parameters
		\STATE $a_t \gets \{ s_i \mid s_i \in M, |s_i| \geq t \}$ \COMMENT{Subjects with $\geq t$ attributes}
		\STATE $r \gets \infty$
		\FORALL{subject $s_i \in a_t$}
		\STATE $\mathcal{SS}_{s_i} \gets \mathsf{ConstructSubjectSpace}(s_i, M, R)$ \COMMENT{Using Alg.~\ref{alg:credential-verification}}
		\STATE $r \gets \min(r, |\mathcal{SS}_{s_i}|)$ \COMMENT{Find minimum subject space size}
		\ENDFOR
		\RETURN $(a_t, r)$
	\end{algorithmic}
\end{algorithm}

Algorithm~\ref{alg:subject-anonymity} calculates the overall anonymity score $\mathcal{A}_{sub}(s)$ for a specific subject $s$. It aggregates the request anonymity values $\mathcal{E}_{req}$ for all possible credentials that $s$ can generate, weighted by the frequency of each credential. This provides a comprehensive measure of the subject's anonymity across all potential requests. The algorithm leverages Algorithms~\ref{alg:rt-anonymity} and~\ref{alg:request-anonymity} to compute the necessary values. This score is crucial for monitoring and enforcing \textbf{G1} at the subject level. It also aids in achieving \textbf{G4} by identifying subjects with low anonymity scores, who may be vulnerable to correlation attacks.
\begin{algorithm}[t]
	\caption{Subject Anonymity Computation}
	\label{alg:subject-anonymity}
	\begin{algorithmic}[1]
		\REQUIRE Subject $s$, subject attribute matrix $M$, request history $R$
		\ENSURE Subject anonymity score $\mathcal{A}_{sub}(s)$
		\STATE $\mathcal{A}_{sub} \gets 0.0$ \COMMENT{Evaluate all possible credential sizes}
		\FOR{$t \gets 1$ to $|s|$} 
		\STATE $(a_t, r) \gets \mathsf{CalculateRTAnonymity}(M, t)$ \COMMENT{Using Alg.~\ref{alg:rt-anonymity}}
		\STATE $\mathsf{total\_subjects} \gets \sum_{s_j \in a_t} |\mathcal{SS}_{s_j}|$
		\FORALL{subject $s_i \in a_t$}
		\STATE $\mathcal{E}_{req} \gets \mathsf{CalculateRequestAnonymity}(s_i, M, R)$ \COMMENT{Using Alg.~\ref{alg:request-anonymity}}
		\STATE $\mathsf{weight} \gets |\mathcal{SS}_{s_i}| / \mathsf{total\_subjects}$
		\STATE $\mathcal{A}_{sub} \gets \mathcal{A}_{sub} + \mathcal{E}_{req} \cdot \mathsf{weight}$ \COMMENT{Weighted sum}
		\ENDFOR
		\ENDFOR
		\RETURN $\mathcal{A}_{sub}$
	\end{algorithmic}
\end{algorithm}

\subsection{Dynamic Optimization Module}
\label{subsec:dyn_opt_design}

The Dynamic Optimization Module adapts attribute weights based on both authorization patterns and anonymity considerations, creating a feedback loop that continuously improves system performance. It plays a key role in achieving \textbf{G3} and \textbf{G4} by optimizing policy evaluation efficiency while maintaining privacy.

\subsubsection{Theoretical Foundation}

\ZJ{Information gain from authorization decisions is combined} with anonymity metrics to compute dynamic attribute weights. The information gain $I(D, A)$ for an attribute $A$ measures how much it reduces uncertainty about authorization outcomes:
\begin{equation}
	I(D, A) = H(D) - H(D|A)
\end{equation}
where $H(D)$ is the entropy of authorization decisions:
\begin{equation}
	H(D) = -\sum_{d \in \{\mathsf{grant}, \mathsf{deny}\}} P(d) \log_2 P(d)
\end{equation}
and $H(D|A)$ is the conditional entropy:
\begin{equation}
	\begin{split}
		H(D|A) = &-\sum_{v \in V_A} P(A=v) \times\\
		&\sum_{d \in \{\mathsf{grant}, \mathsf{deny}\}}
		P(d|A=v) \log_2 P(d|A=v)
	\end{split}
\end{equation}
\ZJ{The final weight combines information gain with the attribute's \textbf{individual anonymity contribution}, which is derived from the $(r,t)$-anonymity assessment (Algorithm~\ref{alg:rt-anonymity}) with $t=1$:
\begin{equation}
	A_w = I(D, A) + \mathcal{A}_{attr}(a_i)
\end{equation}
where $\mathcal{A}_{attr}(a_i)$ represents the anonymity contribution of attribute $a_i$ when considered individually, computed as the minimum credential subject space size $r$ from Algorithm~\ref{alg:rt-anonymity} with $t=1$.
}

\subsubsection{Algorithm Implementation}
Algorithm~\ref{alg:context-entropy} computes the information gain $I(D, a)$ for a given attribute $a$ based on the \ac{AAH} Pool. It calculates the reduction in uncertainty about authorization decisions when the value of attribute $a$ is known. This algorithm supports \textbf{G3} by identifying attributes that are most predictive of access outcomes, allowing for efficient policy structuring. It also contributes to \textbf{G4} by ensuring that attributes with high information gain are prioritized in the policy tree, reducing the risk of attribute correlation attacks by minimizing the number of attributes needed for decisions.

\begin{algorithm}[t]
	\caption{Context-Aware Entropy Calculation}
	\label{alg:context-entropy}
	\begin{algorithmic}[1]
		\REQUIRE \ac{AAH} Pool $H$, attribute $a$
		\ENSURE Information gain $I(D, a)$
		\STATE $H(D) \gets -\sum_{d \in \{\mathsf{grant}, \mathsf{deny}\}} P(d) \log_2 P(d)$
		\STATE $H(D|a) \gets 0.0$
		\FORALL{value $v \in \mathsf{domain}(a)$}
		\STATE $P(v) \gets \mathsf{frequency\ of\ } v \mathsf{\ in\ } H$
		\STATE $H(D|a) \gets H(D|a) + P(v) \cdot H(D | a = v)$
		\ENDFOR
		\STATE $I(D, a) \gets H(D) - H(D|a)$
		\RETURN $I(D, a)$
	\end{algorithmic}
\end{algorithm}

Algorithm~\ref{alg:weight-optimization} generates a sorted list of attribute weights by combining the information gain from Algorithm~\ref{alg:context-entropy} with the anonymity scores from the \ac{AQC}. The combined weight $w_i$ reflects both the attribute's decision-making power and its privacy impact. This algorithm is central to \textbf{G3}, as the weight list directly guides the construction of the \ac{EWPT}, ensuring that the most discriminative and privacy-preserving attributes are checked first. It also enhances \textbf{G4} by dynamically adjusting weights to mitigate correlation risks based on current access patterns.
\ZJ{
This continuous, data-driven update mechanism enables \sys{} to naturally adapt to evolving policies. When policies are added, removed, or modified, the resulting changes in access patterns are captured in the \ac{AAH} Pool. Subsequent weight recalculations and \ac{EWPT} reconstructions automatically incorporate these changes, ensuring that the authorization structure remains optimized for the current policy environment without manual intervention.
}
\begin{algorithm}[t]
	\caption{Attribute Weight Optimization}
	\label{alg:weight-optimization}
	\begin{algorithmic}[1]
		\REQUIRE Attribute space $A$, \ac{AAH} Pool $H$
		\ENSURE Sorted attribute weight list $W$
		\STATE $W \gets \emptyset$
		\FORALL{attribute $a_i \in A$}
		\STATE $I(D, a_i) \gets \mathsf{CalculateInformationGain}(H, a_i)$ \COMMENT{Using Alg.~\ref{alg:context-entropy}}
		\ZJ{
			\STATE $(a_1, r) \gets \mathsf{CalculateRTAnonymity}(M, 1)$ \COMMENT{Call Alg.~\ref{alg:rt-anonymity} with $t=1$}
			\STATE $\mathcal{A}_{attr}(a_i) \gets r$ \COMMENT{Use min subject space size as anonymity measure}
		}
		\STATE $w_i \gets I(D, a_i) + \ZJ{\mathcal{A}_{attr}(a_i)}$ \COMMENT{Combined metric}
		\STATE $W \gets W \cup \{(a_i, w_i)\}$
		\ENDFOR
		\STATE $\mathsf{sort}(W)$ by $w_i$ descending
		\RETURN $W$
	\end{algorithmic}
\end{algorithm}

\subsection{Access Control Module}
\label{subsec:ac_module_design}

The Access Control Module implements efficient policy evaluation through the \ac{EWPT} structure, incorporating cryptographic verification for security and path-based matching for efficiency. It is responsible for achieving \textbf{G3} by enabling fast authorization decisions and upholding \textbf{G1} and \textbf{G2} through integrated anonymity checks. 

\begin{figure}[t]
	\centering
	\includegraphics[width=0.5\textwidth]{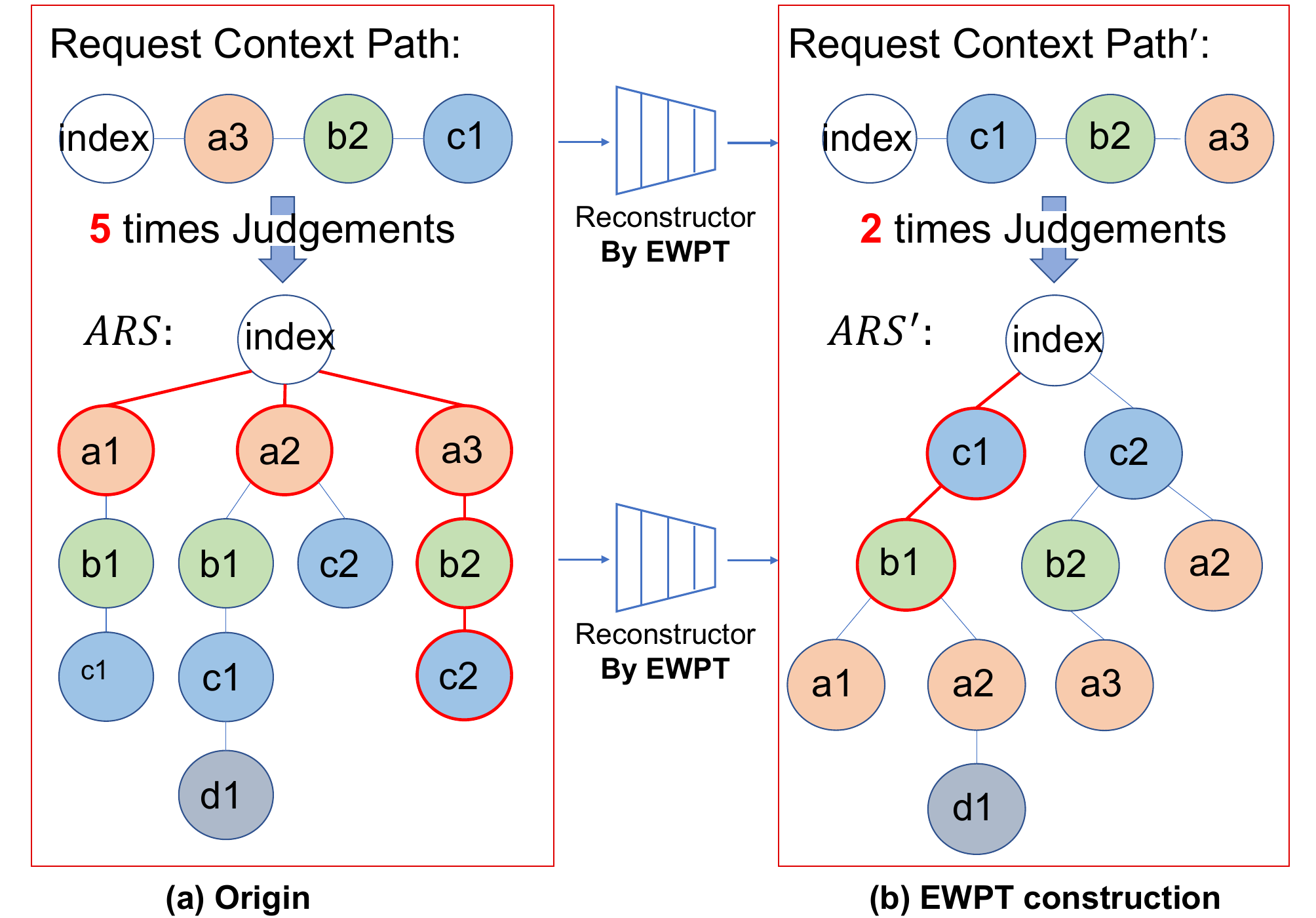}
	\caption{Entropy-Weighted Path Tree structure: (a) initial policy organization; (b) optimized structure after dynamic weight adjustment. The tree structure enables $O(m)$ authorization time complexity where $m$ is the number of attributes.}
	\label{fig:ewpt_structure}
\end{figure}

Algorithm~\ref{alg:ewpt-construction} builds the \ac{EWPT} using the policy rule set and the attribute weight list from Algorithm~\ref{alg:weight-optimization}. The tree is constructed by sorting attributes in each rule by their weight descending, ensuring that high-weight attributes (those with high information gain and anonymity) form the shared prefixes of paths. This structure dramatically reduces the average depth of policy matching, enabling $O(m)$ time complexity where $m$ is the number of attributes in a request. 
To facilitate comprehension, Fig.\ref{fig:ewpt_structure}a shows a simplified example of $P_{W1}$ with the initial attribute weight $W_1 = [a, b, c, d]$, containing four access policy rules: $P = \{(a1, b1, c1), (a2, b1, c1, d1), (a2, c2), (a3, b2, c2)\}$. 
When the attribute weights are updated to $W_2 = [c, b, a, d]$, the policy $P$ is reconstructed into $P_{W2}$, as shown in Fig.\ref{fig:ewpt_structure}b.
For an anonymous request $R = (a3, b2, c1)$, the tree-based matching of origin $P_{W1}$ requires $5$ attempts, while the \ac{EWPT} $P_{W2}$ further optimizes the process, needing only $2$ comparisons. 
This algorithm is pivotal for \textbf{G3}, as it optimizes the policy structure for efficient evaluation. The tree's design also supports \textbf{G4} by promoting the use of attributes that are less susceptible to correlation.

\begin{algorithm}[t]
	\caption{EWPT Construction}
	\label{alg:ewpt-construction}
	\begin{algorithmic}[1]
		\REQUIRE Policy rule set $P$, attribute weight list $W$
		\ENSURE Entropy-Weighted Path Tree $T$
		\STATE $T \gets \mathsf{CreateRootNode}()$
		\FORALL{rule $r \in P$}
		\STATE $\mathsf{sorted\_attrs} \gets \mathsf{sort}(r.\mathsf{attributes})$ by weight in $W$ descending
		\STATE $\mathsf{current} \gets T.\mathsf{root}$
		\FORALL{attribute $a_j \in \mathsf{sorted\_attrs}$}
		\IF{$a_j \notin \mathsf{current}.\mathsf{children}$}
		\STATE $\mathsf{current}.\mathsf{children}[a_j] \gets \mathsf{CreateNode}(a_j)$
		\ENDIF
		\STATE $\mathsf{current} \gets \mathsf{current}.\mathsf{children}[a_j]$
		\ENDFOR
		\STATE $\mathsf{MarkAsLeaf}(\mathsf{current})$ \COMMENT{Complete policy path}
		\ENDFOR
		\RETURN $T$
	\end{algorithmic}
\end{algorithm}

Algorithm~\ref{alg:authorization} is the culmination of the \sys{} framework, performing the final authorization decision. It integrates cryptographic verification, anonymity validation, and policy path checking in a three-step process. First, it verifies the digital signature on the credential to ensure authenticity (using the same method as Algorithm~\ref{alg:credential-verification}). Second, it checks the request's anonymity score against a threshold to ensure sufficient privacy (using Algorithm~\ref{alg:request-anonymity}). Finally, it traverses the \ac{EWPT} to check for a matching path. This algorithm directly enforces \textbf{G1} and \textbf{G2} by rejecting requests that fail anonymity or verification checks. It achieves \textbf{G3} through efficient path matching and contributes to \textbf{G4} by ensuring that only requests with safe attribute combinations are granted.
\begin{algorithm}[t]
	\caption{Authorization Decision with Cryptographic Verification}
	\label{alg:authorization}
	\begin{algorithmic}[1]
		\REQUIRE Request $req = (\sigma_s(c), o, op, env)$, \ac{EWPT} $T$, weight list $W$, public key $pk_s$
		\ENSURE Authorization decision: $\mathsf{GRANT}$ or $\mathsf{DENY}$
		\STATE \textbf{Step 1: Cryptographic Verification}
		\STATE $\mathsf{valid} \gets \mathsf{Verify}(pk_s, c, \sigma_s(c))$
		\IF{not $\mathsf{valid}$}
		\RETURN $\mathsf{DENY}$ \COMMENT{Reject unverifiable request}
		\ENDIF
		
		\STATE \textbf{Step 2: Anonymity Validation}
		\STATE $\mathcal{E}_{req} \gets \mathsf{CalculateRequestAnonymity}(req)$ \COMMENT{Using Alg.~\ref{alg:request-anonymity}}
		\IF{$\mathcal{E}_{req} < \mathsf{threshold}$}
		\RETURN $\mathsf{DENY}$ \COMMENT{Insufficient anonymity}
		\ENDIF
		
		\STATE \textbf{Step 3: Policy Path Evaluation}
		\STATE $\mathsf{attr\_sequence} \gets \mathsf{ExtractAndSortAttributes}(req, W)$
		\STATE $\mathsf{current} \gets T.\mathsf{root}$
		\FORALL{value $v_i \in \mathsf{attr\_sequence}$}
		\IF{$v_i \notin \mathsf{current}.\mathsf{children}$}
		\RETURN $\mathsf{DENY}$ \COMMENT{No matching path}
		\ENDIF
		\STATE $\mathsf{current} \gets \mathsf{current}.\mathsf{children}[v_i]$
		\ENDFOR
		
		\IF{$\mathsf{current}.\mathsf{isLeaf}$}
		\RETURN $\mathsf{GRANT}$ \COMMENT{Valid path exists}
		\ELSE
		\RETURN $\mathsf{DENY}$ \COMMENT{Incomplete path}
		\ENDIF
	\end{algorithmic}
\end{algorithm}

The integrated design of these eight algorithms creates a comprehensive framework that provides both strong privacy guarantees through information-theoretic anonymity metrics and efficient authorization through optimized policy structures, all while maintaining cryptographic security through digital signature verification, forming a framework that achieves the goals of subject anonymity, unlinkability, efficient access control, and resilience to attribute correlation attacks.

%% file: 06.Security.tex
\section{Security Analysis}
\label{sec:security_analysis}

This section provides a formal reduction-based security analysis of the \sys{} framework. \ZJ{It is demonstrated} that breaking the anonymity of \sys{} is computationally equivalent to solving well-established hard problems under the defined threat model (Section~\ref{subsec:threat_model}).

\subsection{Assumptions and Security Analysis}

The security reduction relies on the standard \ac{DL} assumption (Assumption~\ref{ass:dl}) and the \ac{EUF-CMA} security of digital signature scheme. \ZJ{The main security theorem is formally stated.}

\begin{theorem}[Security of \sys{}]
	\label{thm:main-security}
	Let $\lambda$ be the security parameter. Let $\mathcal{A}$ be any \ac{PPT} adversary against the \textit{Request Anonymity} (Def.~\ref{def:request_anonymity}) or \textit{Request Unlinkability} (Def.~\ref{def:request_unlinkability}) of the \sys{} framwork $\Pi$, with advantage $\mathsf{Adv}_{\mathcal{A},\Pi}^{\mathsf{Anon}}(\lambda)$.
	If the digital signature scheme $\Sigma$ is \ac{EUF-CMA} secure and the \ac{DL} assumption holds in group $\mathbb{G}$, then $\mathsf{Adv}_{\mathcal{A},\Pi}^{\mathsf{Anon}}(\lambda)$ is negligible.
	Formally, there exists \ac{PPT} algorithms (simulators) $\mathcal{S}_1$ and $\mathcal{S}_2$ such that:
	\begin{equation}
		\mathsf{Adv}_{\mathcal{A},\Pi}^{\mathsf{Anon}}(\lambda) \leq \mathsf{Adv}_{\Sigma}^{\mathsf{EUF-CMA}}(\mathcal{S}_1(\mathcal{A})) + \mathsf{Adv}_{\mathbb{G}}^{\mathsf{DL}}(\mathcal{S}_2(\mathcal{A})) + \mathsf{negl}(\lambda)
	\end{equation}
\end{theorem}

The proof of Theorem~\ref{thm:main-security} is structured as a sequence of games. The core argument is a reduction showing that any successful anonymity adversary $\mathcal{A}$ can be used to either break the \ac{EUF-CMA} security of the signature scheme or to compute discrete logarithms. \ZJ{A detailed proof sketch is provided here,} with the full formal proof deferred to Appendix~\ref{app:security-proof}. \ZJ{It is} first argued that the signature scheme ensures the \textit{authenticity} and \textit{integrity} of the attribute credential $c$. \ZJ{This implies} that the credential subject space $\mathcal{SS}_c$ is constructed from a valid, unaltered credential that was indeed signed by a registered subject.

Given this, the entropy $\mathcal{E}_{req}(req_c)$ becomes a function of the legitimate system parameters ($S$, $R$) and the credential $c$. The adversary $\mathcal{A}$'s advantage in the anonymity game must therefore stem from an ability to link the signature $\sigma_s(c)$ to the private key $sk_s$, even when the credential $c$ itself is non-unique. \ZJ{Two cases are distinguished:}

\begin{enumerate}
	\item \textbf{Case 1: Forgery.} If the adversary can produce a valid request $req^*$ with a forged signature $\sigma^*$ on a credential $c^*$ that has not been signed by the claimed subject, this directly contradicts the \ac{EUF-CMA} security of the signature scheme. A simulator $\mathcal{S}_1$ can use $\mathcal{A}$ to forge a signature, winning the \ac{EUF-CMA} game.
	
	\item \textbf{Case 2: Extraction.} If the adversary wins without a forgery, it must be leveraging the signature itself to gain information about the signer's identity, beyond what is revealed by $c$. For example, if the signature is not perfectly opaque (e.g., if randomness is reused in \ac{ECDSA}), it might leak information about the private key. (\ZJ{A discussion on the signature selection is provided} in Section ~\ref{subsec:signature-choice}) A simulator $\mathcal{S}_2$ can embed a \ac{DL} challenge $h = g^a$ into the public key of the challenge subject. If $\mathcal{A}$ can distinguish which subject signed a credential, $\mathcal{S}_2$ can potentially extract the discrete logarithm $a$ from the adversary's behavior.
\end{enumerate}

The full proof in Appendix~\ref{app:security-proof} formalizes this intuition, constructs the simulators $\mathcal{S}_1$ and $\mathcal{S}_2$ in detail, and provides the probability calculation showing that the advantage of $\mathcal{A}$ is bounded by the advantages of the simulators in solving the underlying problems.

\subsection{Discussion on the Security of Signature Schemes}
\label{subsec:signature-choice}

The reduction in Case 2 is most straightforward if the signature scheme is \textit{deterministic} (e.g., a deterministic variant of EdDSA \cite{EdDSA}) or if it can be modeled as a random oracle. Deterministic signatures ensure that the same credential $c$ always produces the same signature $\sigma_s(c)$, eliminating the possibility that randomness in the signature itself provides additional information to the adversary. The reduction proof in the Appendix~\ref{app-signature} assumes a deterministic signature scheme for clarity. The security holds for probabilistic schemes under the \ac{ROM}.

%% file: 07.Performance.tex
\section{Performance Evaluation}
\label{sec:evaluation}

This section presents a comprehensive empirical analysis of the proposed \sys{} framework. To rigorously evaluate its effectiveness, \ZJ{the following three \acp{RQ} are addressed:}

\begin{description}
    \item[\textbf{RQ1:}]  How do different system parameters (e.g., number of subjects, attributes) affect the anonymity guarantees provided by \sys{}'s quantification module?
    \item[\textbf{RQ2:}] Does \sys{} achieve significant performance improvements over state-of-the-art baselines? If so, what is the contribution of each key innovation (\ac{EWPT} structure vs. dynamic optimization)?
    \item[\textbf{RQ3:}] How does \sys{} perform under increasing system scale and complexity? Is it resilient to the "attribute explosion" problem?
\end{description}

\ZJ{The experimental setup is detailed, the design of the test cases is described, and a rigorous evaluation is presented} to answer these \ac{RQ}.

\subsection{Experimental Setup}
\label{subsec:experimental_setup}

To thoroughly assess the performance of \sys{}, \ZJ{a simulation testbed was established} on an Apple M1 Pro platform (16 GB RAM) running macOS Monterey v12.3. The environment was built on Hyperledger Fabric v2.2.1, with containerization managed by Docker v20.10.13 and Docker-compose v1.29.2. The smart contracts (chaincode) for the \ac{AQC}, \ac{AIC}, and \ac{PDC}/\ac{PEC}/\ac{PAC} modules were developed in Golang v1.18 and deployed onto the Fabric network. \ZJ{The EdDSA signature scheme was employed} using the Ed25519 curve, providing a security level of $\lambda=128$ bits with SHA-256 as the hash function. To ensure the reliability and stability of the results, each experiment was repeated 10 times, and the average values are reported for analysis.

\subsection{Test Case Design}
\label{subsec:test_case_design}

To answer \textbf{RQ1} and \textbf{RQ3}, \ZJ{\sys{} must be evaluated} under a wide range of configurations. \ZJ{Test cases were constructed} based on features extracted from a real-world IoT dataset~\cite{Ahmed2024DatasetFA}, which focuses on physical-layer feature-based authentication and authorization. This dataset, collected via Zigbee Zolertia Z1 nodes, provides a solid foundation for simulating realistic access control scenarios. \ZJ{Synthetic yet representative data} for subjects, objects, access requests, and policies \ZJ{were generated} based on the fields and entries in this dataset.

Guided by the \ac{OVAT} principle~\cite{azmi2018optimization}, \ZJ{seven key influencing factors were identified and 15 test cases were designed.} The core variables include the number of entities (subjects, objects), the volume of actions (requests), the complexity of control (policies), and the dimensionality of attributes (value range, subject/object attribute counts). The specific parameters for all 15 test cases (C1--C15) are meticulously detailed in Table~\ref{table:test_cases}. The scales for the number of subjects (5K--15K) and requests (500K--1500K) were chosen to closely mirror the potential scales encountered in real-world applications like smart manufacturing and healthcare, ensuring the practical relevance of this evaluation.

\begin{table}[h]
	\centering
	\caption{Parameters for the Designed Test Cases}
	\label{table:test_cases}
	\resizebox{0.5\textwidth}{!}{ 
		\begin{tabular}{ccccccccc}
			\toprule
			\textbf{\makecell[c]{Test \\ Case}} & \textbf{\makecell[c]{\# Sub- \\jects}} & \textbf{\makecell[c]{\# Ob- \\jects}} & \textbf{\makecell[c]{\# Re- \\ quests}} & \textbf{\makecell[c]{\# Po- \\licies}} & \textbf{\makecell[c]{Attr. \\ Value \\ Range}} & \textbf{\makecell[c]{\# Sub. \\ Attrs.}} & \textbf{\makecell[c]{\# Obj. \\ Attrs.}} \\
			\midrule
			C1  & \textbf{5K} & 10K & 1000K & 100 & 4 & 4 & 2 \\
			C2  & 10K & 10K & 1000K & 100 & 4 & 4 & 2 \\
			C3  & \textbf{15K} & 10K & 1000K & 100 & 4 & 4 & 2 \\
			C4  & 10K & \textbf{5K} & 1000K & 100 & 4 & 4 & 2 \\
			C5  & 10K & \textbf{15K} & 1000K & 100 & 4 & 4 & 2 \\
			C6  & 10K & 10K & \textbf{500K} & 100 & 4 & 4 & 2 \\
			C7  & 10K & 10K & \textbf{1500K} & 100 & 4 & 4 & 2 \\
			C8  & 10K & 10K & 1000K & \textbf{50} & 4 & 4 & 2 \\
			C9  & 10K & 10K & 1000K & \textbf{150} & 4 & 4 & 2 \\
			C10 & 10K & 10K & 1000K & 100 & \textbf{2} & 4 & 2 \\
			C11 & 10K & 10K & 1000K & 100 & \textbf{6} & 4 & 2 \\
			C12 & \textbf{15K} & 10K & 1000K & 100 & 4 & \textbf{5} & 2 \\
			C13 & \textbf{15K} & 10K & 1000K & 100 & 4 & \textbf{3} & 2 \\
			C14 & 10K & 10K & 1000K & 100 & \textbf{2} & 4 & \textbf{4} \\
			C15 & 10K & 10K & 1000K & 100 & \textbf{2} & 4 & \textbf{3} \\
			\bottomrule
		\end{tabular}
	}
\end{table}

The test cases are grouped to analyze the impact of each factor, which is crucial for answering RQ1 and RQ3:
\begin{itemize}[noitemsep, topsep=0pt]
	\item \textbf{Subject Quantity (C1, C2, C3):} Tests system scalability w.r.t. user base size.
	\item \textbf{Object Quantity (C2, C4, C5):} Tests scalability w.r.t. resource base size.
	\item \textbf{Request Quantity (C2, C6, C7):} Tests performance under different load intensities.
	\item \textbf{Policy Quantity (C2, C8, C9):} Tests resilience to growing policy complexity.
	\item \textbf{Attribute Value Range (C2, C10, C11):} Tests impact of attribute granularity.
	\item \textbf{Subject Attribute Quantity (C3, C12, C13):} Tests resilience to subject attribute explosion.
	\item \textbf{Object Attribute Quantity (C10, C14, C15):} Tests resilience to object attribute explosion.
\end{itemize}
This structured design ensures a controlled and comparable basis for evaluating the impact of each variable. 
\ZJ{
\textbf{Note} that the core innovation of \sys{} breaks the traditional privacy-efficiency trade-off. As demonstrated in Section \ref{subsec:anonymity_analysis} and Section \ref{subsec:efficiency_analysis}, \sys{} maintains high anonymity under diverse conditions while achieving significant performance gains. This proves that \sys{} not merely balances these dual objectives, but simultaneously delivers substantial improvements along both dimensions.
}

\subsection{Anonymity Analysis}
\label{subsec:anonymity_analysis}

This subsection addresses \textbf{RQ1} by evaluating the capability of \sys{} to preserve subject unlinkability under fine-grained policies. The analysis was conducted in two steps: 1) assessing the request anonymity distribution across all possible $(r, t)$-pairs, and 2) quantifying the overall anonymity of each subject based on the resulting distribution.

\subsubsection{Graphical Interpretation}
The results for each influencing factor group are presented in paired figures (e.g., Fig.~\ref{fig:anon_subject_count}). For each group:
\begin{itemize}[noitemsep, topsep=0pt]
	\item The \textbf{left subfigure} (Fig.~\ref{fig:anon_subject_count}a) shows the request anonymity distribution across different $(r, t)$-pairs. The x-axis represents the test cases, the y-axis represents the request anonymity ($\mathcal{E}_{req}$), and different-colored legends represent different $t$ values. Comparisons between bars reveal the factor's impact, while comparisons within legends show the effect of varying $t$.
	\item The \textbf{right subfigure} (Fig.~\ref{fig:anon_subject_count}b) displays the corresponding subject anonymity distribution ($\mathcal{A}_{sub}$) for the same test case group, presented using box plots to illustrate the central tendency and dispersion of anonymity scores across the subject population.
\end{itemize}

\subsubsection{Key Findings (Answering RQ1)}
The analysis of all figure pairs led to the following conclusions, which directly characterize the anonymity properties of \sys{}:

\begin{figure}[h]
	\centering
	\includegraphics[width=0.5\textwidth]{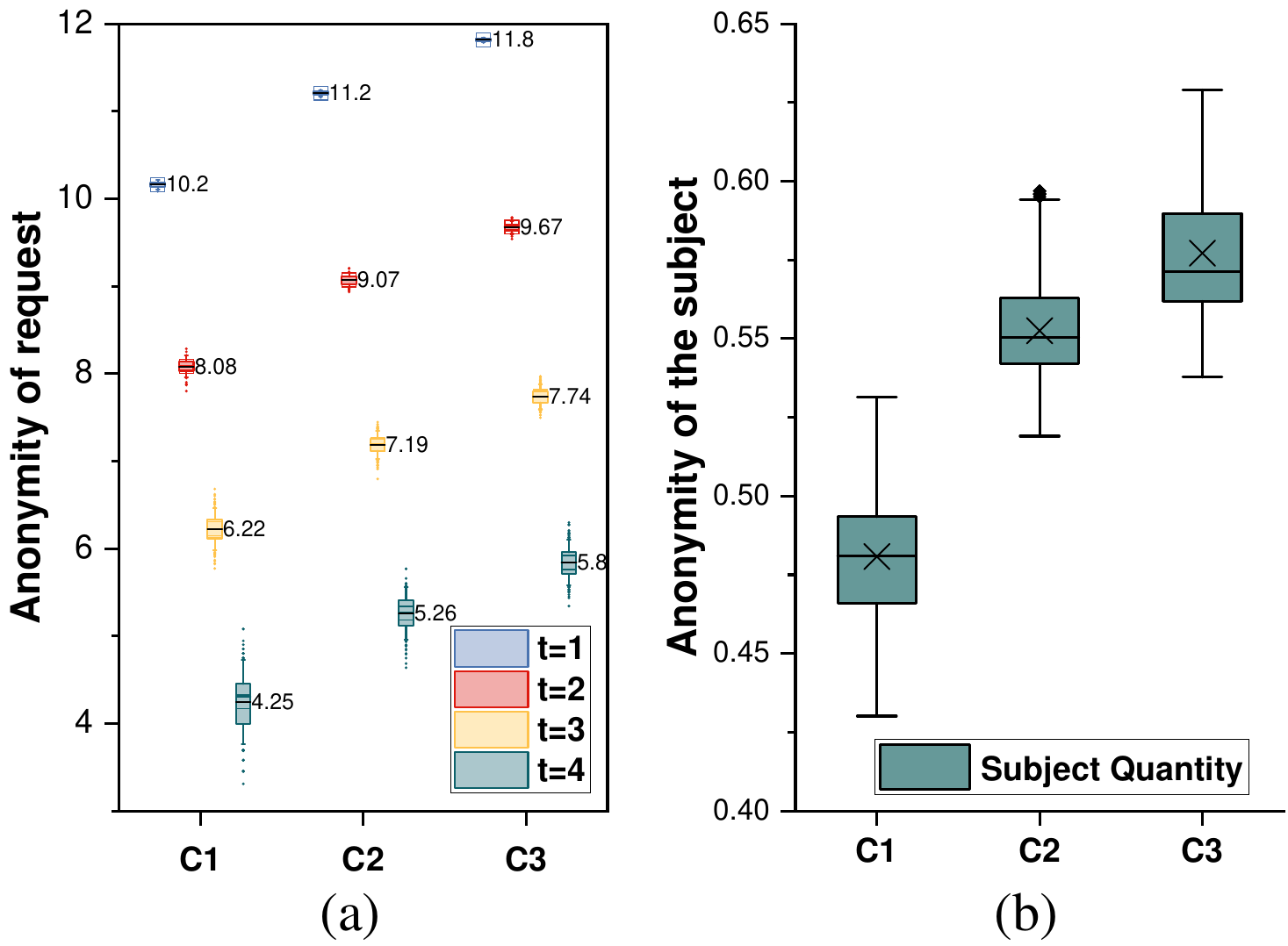}
	\caption{\ZJ{The Impact of \textbf{Subject Quantity} on Anonymity Distribution}}
	\label{fig:anon_subject_count}
\end{figure}

\begin{figure}[h]
	\centering
	\includegraphics[width=0.5\textwidth]{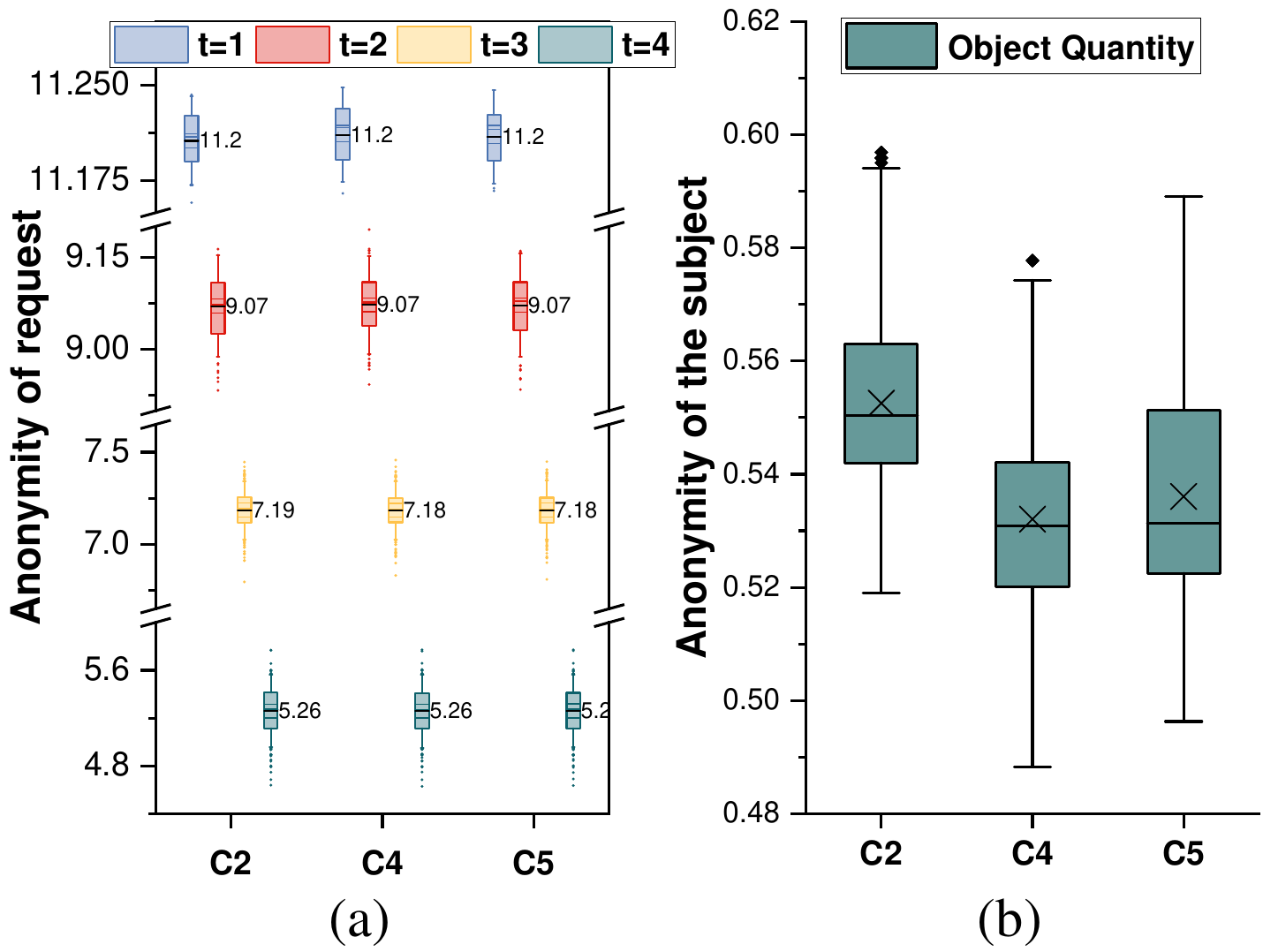}
	\caption{\ZJ{The Impact of \textbf{Object Quantity} on Anonymity Distribution}}
\label{fig:anon_object_count}
\end{figure}

\begin{figure}[h]
	\centering
	\includegraphics[width=0.5\textwidth]{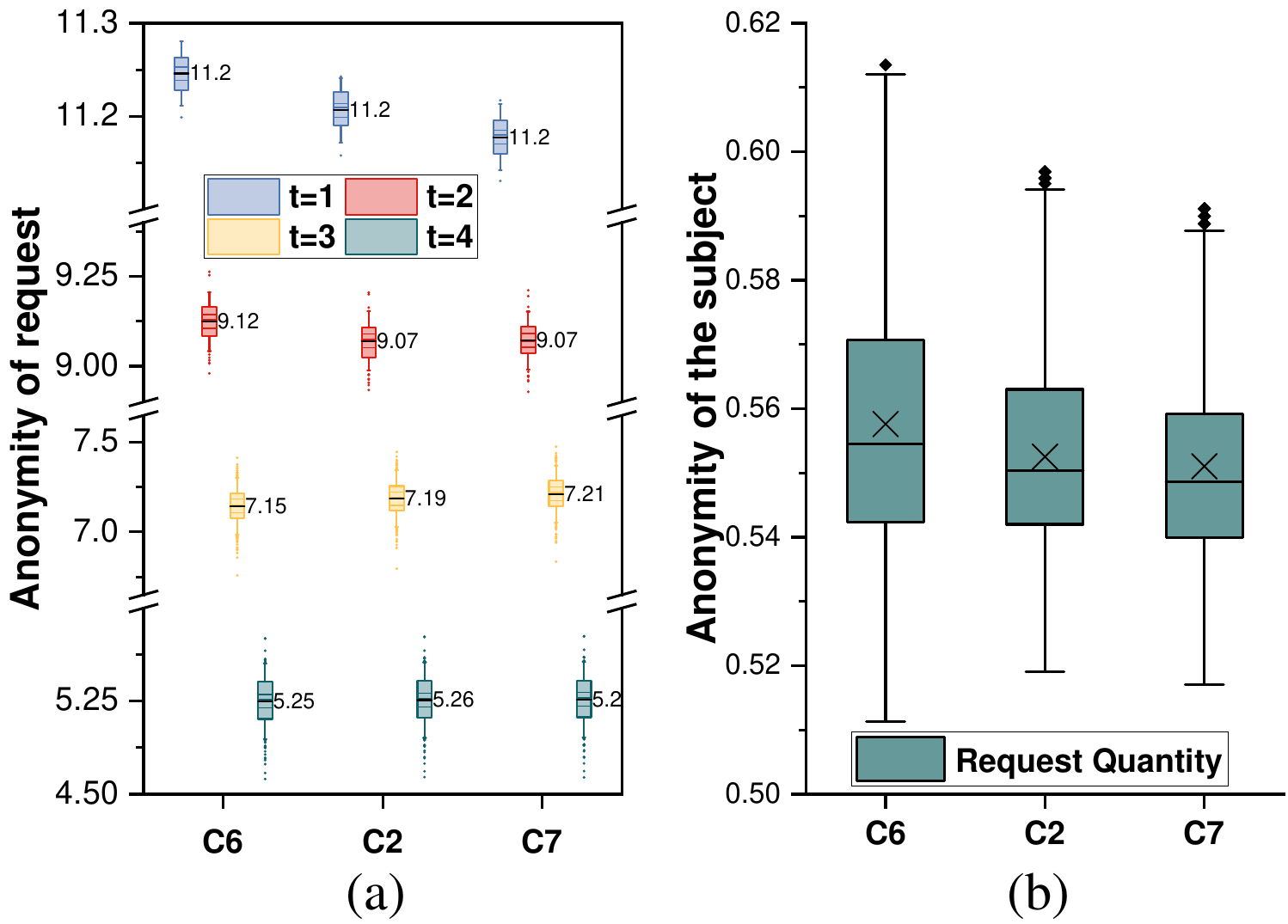}
	\caption{\ZJ{The Impact of \textbf{Request Quantity} on Anonymity Distribution}}
\label{fig:anon_request_count}
\end{figure}

\begin{figure}[h]
	\centering
	\includegraphics[width=0.5\textwidth]{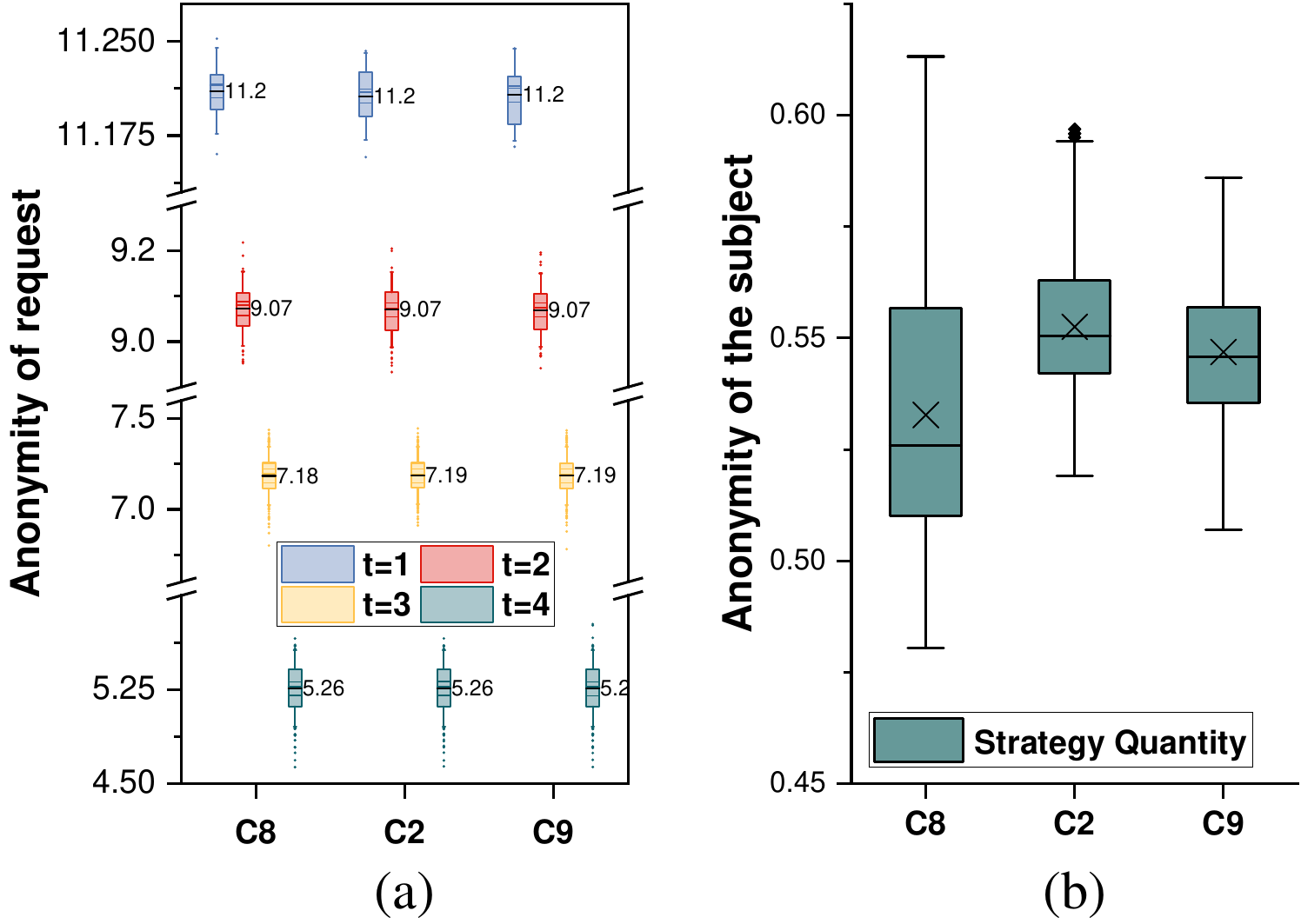}
	\caption{\ZJ{The Impact of \textbf{Strategy Quantity} on Anonymity Distribution}}
\label{fig:anon_policy_count}
\end{figure}

\begin{figure}[h]
	\centering
	\includegraphics[width=0.5\textwidth]{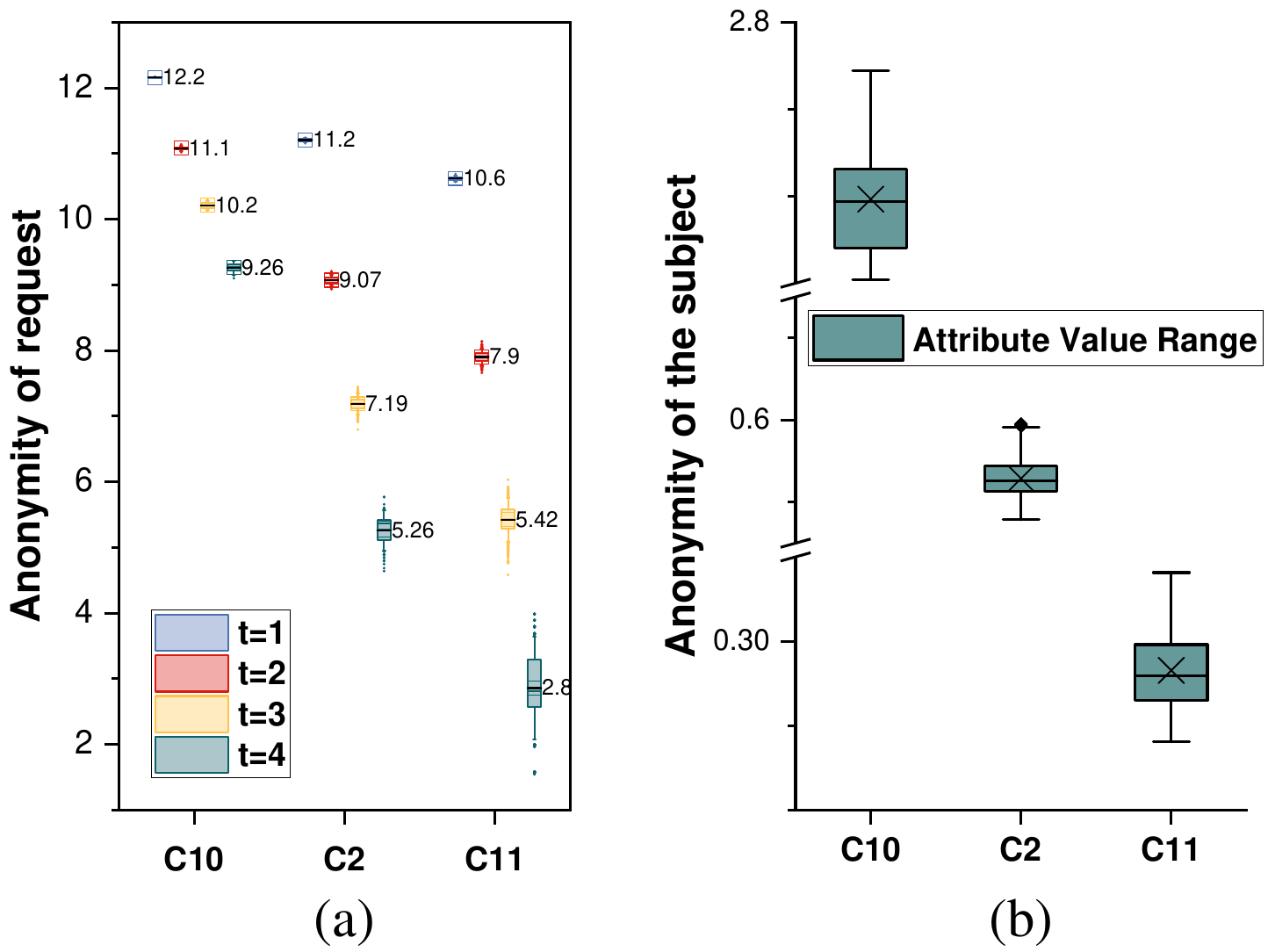}
	\caption{\ZJ{The Impact of \textbf{Attribute Value Range} on Anonymity Distribution}}
\label{fig:anon_attr_value}
\end{figure}

\begin{figure}[h]
	\centering
	\includegraphics[width=0.5\textwidth]{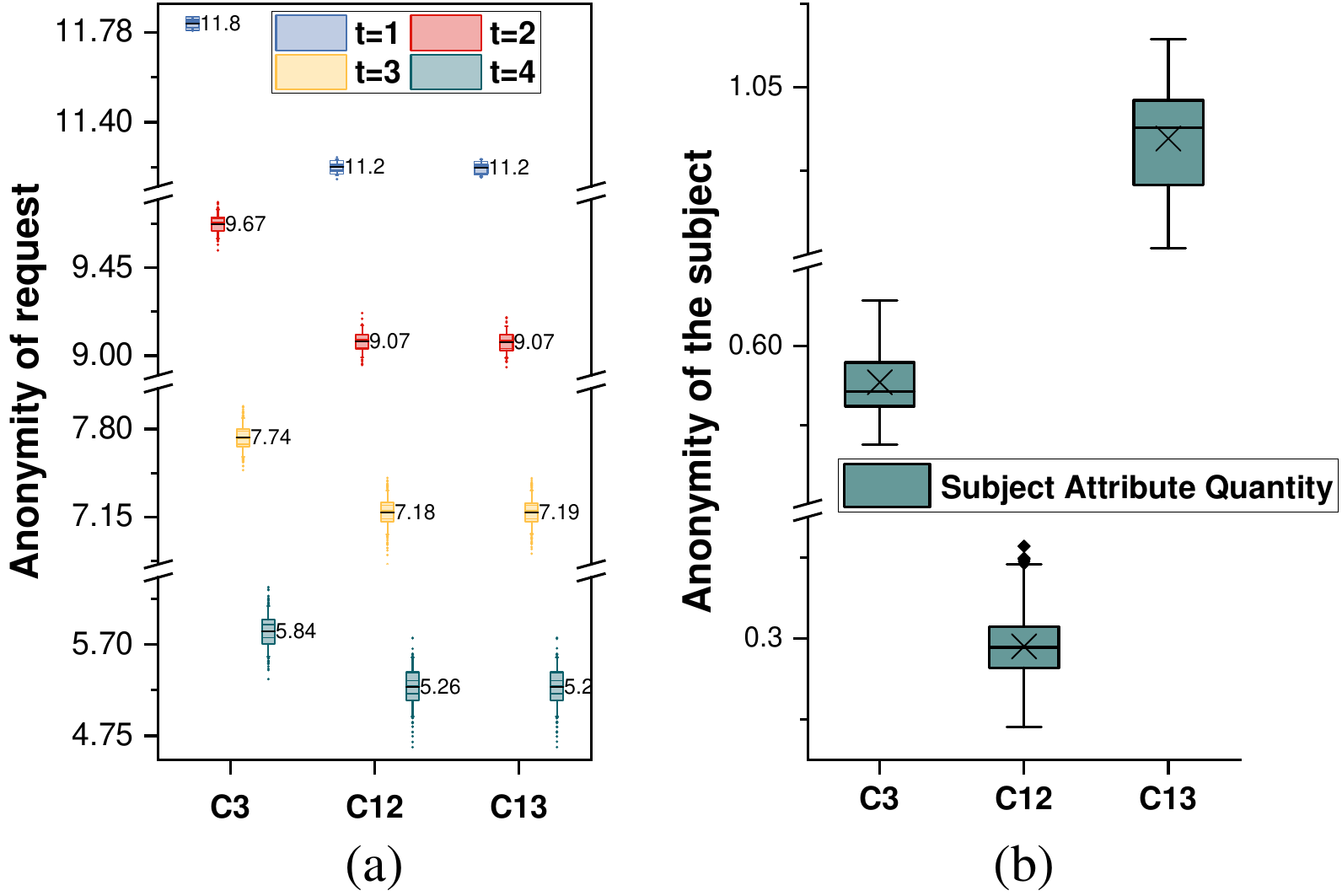}
	\caption{\ZJ{The Impact of \textbf{Subject Attribute Quantity} on Anonymity Distribution}}
\label{fig:anon_sub_attr_count}
\end{figure}

\begin{figure}[h]
	\centering
	\includegraphics[width=0.5\textwidth]{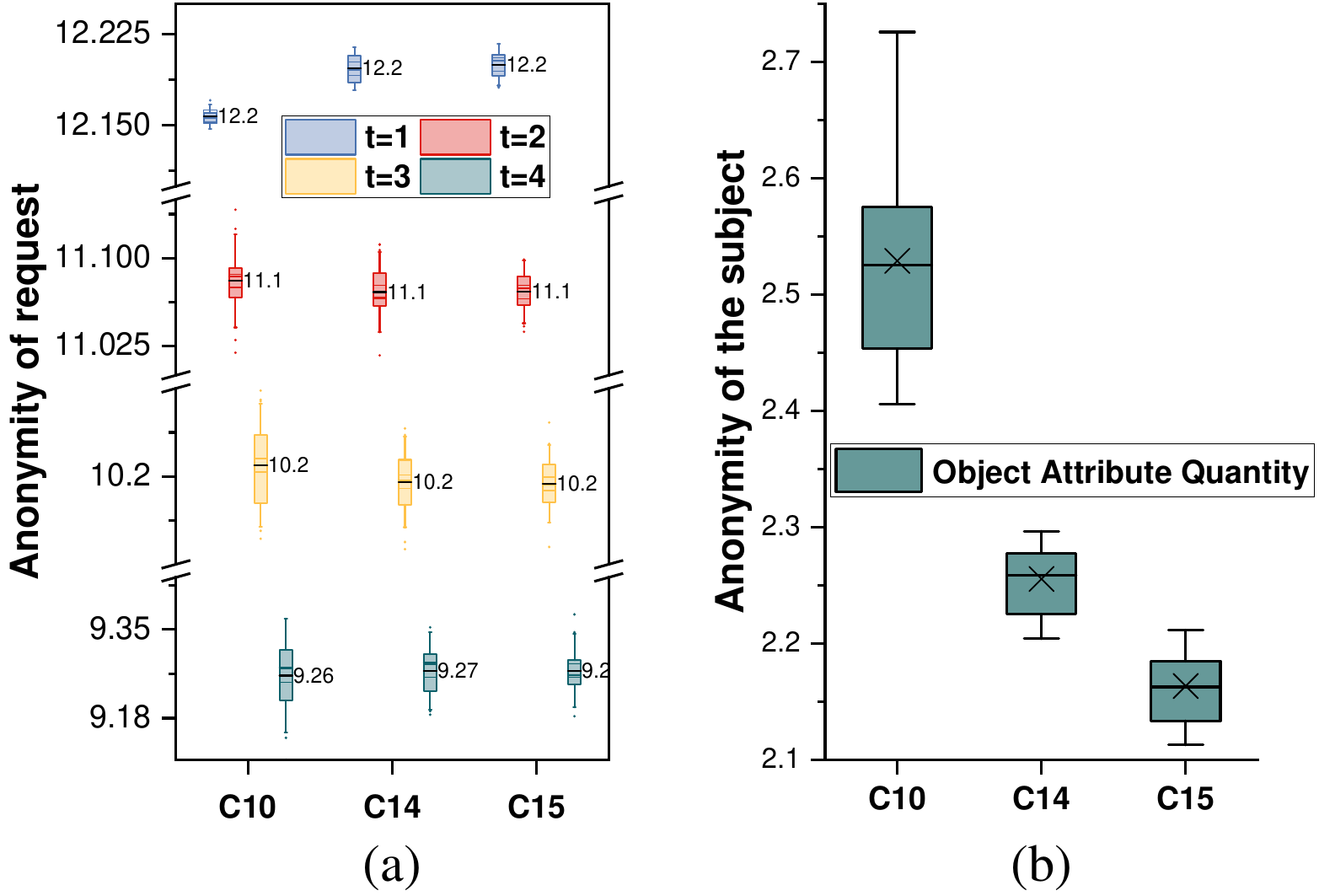}
	\caption{\ZJ{The Impact of \textbf{Object Attribute Quantity} on Anonymity Distribution}}
	\label{fig:anon_obj_attr_count}
\end{figure}

\begin{enumerate}[label=(\roman*), wide]
	\item \textbf{Inverse Correlation with $t$:} A strong negative correlation was observed between the number of non-empty attributes $t$ a subject possesses and the anonymity of requests it can generate (From Fig.~\ref{fig:anon_subject_count}a to Fig.~\ref{fig:anon_obj_attr_count}a). As $t$ decreases, the set $a_t$ expands, leading to a larger credential subject space $\mathcal{SS}_c$ and consequently higher request anonymity. This effect is fundamental and outweighs the impact of other variables like request count. \textbf{This finding confirms that the $(r,t)$-anonymity model effectively captures the fundamental trade-off between the amount of attribute information revealed and the level of anonymity provided.}
	
	\item \textbf{Positive Correlation with Subject Population:} Both request and subject anonymity exhibited a positive correlation with the number of subjects (Fig.~\ref{fig:anon_subject_count}). A larger subject space $S$ increases the potential reuse of attribute combinations, expanding $\mathcal{SS}_c$ and the adversary's uncertainty. The increase in subject anonymity was slightly less pronounced than for requests, suggesting that $\mathcal{A}_{sub}$ might be influenced by factors like individual request frequency. \textbf{This demonstrates that \sys{} is well-suited for large-scale deployments, as its anonymity guarantees strengthen with a growing user base.}
	
	\item \textbf{Indirect Impact of Object/Request/Policy Quantity:} Variations in the number of objects, requests, or policies did not directly alter the structure of attribute credentials but significantly affected the distribution frequency of request types (Fig.~\ref{fig:anon_object_count},~\ref{fig:anon_request_count},~\ref{fig:anon_policy_count}). This, in turn, influenced the composition of the historical subject set $\mathcal{S}_2^c$ within $\mathcal{SS}_c$, affecting the aggregation and, thus, the entropy of the subject distribution. \textbf{This shows that while these factors influence anonymity, the core determinant remains the attribute distribution captured by the credential subject space.}
	
	\item \textbf{Negative Correlation with Attribute Factors:} Factors related to attribute complexity—value range, number of subject attributes, and number of object attributes—showed a clear negative correlation with anonymity (Fig.~\ref{fig:anon_attr_value},~\ref{fig:anon_sub_attr_count},~\ref{fig:anon_obj_attr_count}). Increasing these factors exponentially expands the potential attribute combination space. Within a finite subject space, this reduces the occurrence frequency of any specific combination, thereby decreasing anonymity levels and increasing distribution disparity. \textbf{This highlights the challenge of ``attribute explosion" and underscores the importance of \sys{}'s dynamic optimization module in mitigating its effects on performance (addressed in RQ3).}
\end{enumerate}
\textbf{Note:} These conclusions hold under the prerequisite of a sufficiently large subject space. An overly small subject space risks creating unique attribute combinations (implicit identifiers), which would breach anonymity and make trend analysis meaningless.

\textbf{Summary for RQ1:} The anonymity quantification module of \sys{} provides consistent and measurable privacy guarantees. Anonymity levels are primarily determined by the subject population size and the number of attributes in a credential, while other factors exert an indirect influence. This validates the design of our information-theoretic metrics.

\subsection{Efficiency Analysis}
\label{subsec:efficiency_analysis}
This subsection addresses \textbf{RQ2} and \textbf{RQ3} by evaluating the performance and scalability of \sys{}. 

\subsubsection{Baseline Selection}
To thoroughly evaluate the efficiency and practicality of \sys{},\ZJ{two baseline schemes were selected} for comparison:
\begin{itemize}[wide]
	\item \textbf{Fabric-IoT}~\cite{Liu2020FabriciotAB}: A representative ABAC model implemented on Hyperledger Fabric. It serves as a benchmark for traditional, non-optimized \ac{ABAC} performance within the same blockchain environment.
	\item \textbf{\sys{}-Static:} An \textbf{ablated variant} of \sys{} where the \textbf{dynamic optimization module is removed}. The \ac{EWPT} structure is built using a fixed, initial attribute weight ordering. Comparing with \sys{}-Static helps isolate and quantify the performance contribution of the dynamic weight update mechanism itself, answering RQ2.
\end{itemize}

\subsubsection{Performance Comparison (Answering RQ2 and RQ3)}
\begin{figure*}[t]
	\centering
	\includegraphics[width=\textwidth]{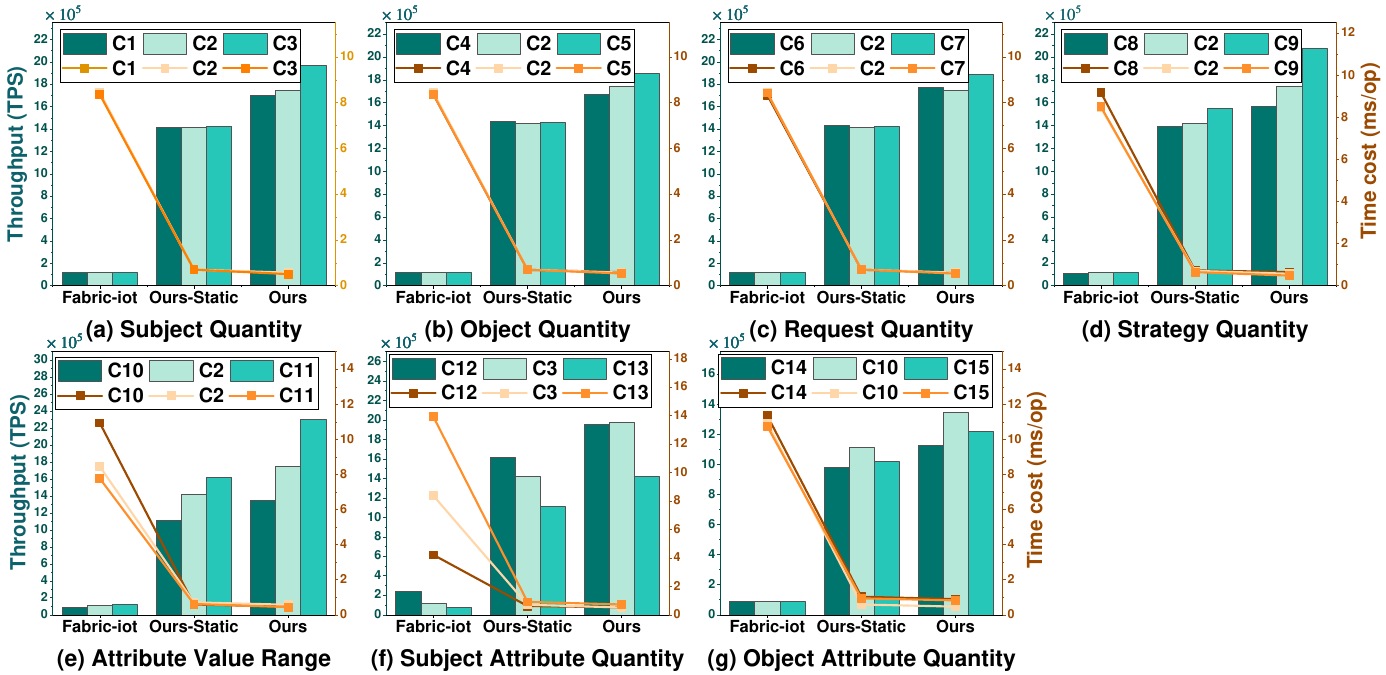}
	\caption{Performance comparison across the seven influencing factor groups. \ZJ{All subfigures (a)-(g) share a common coordinate system: the left y-axis (green) represents \textbf{Throughput (TPS)} and the right y-axis (orange) represents \textbf{Latency (ms)}}. The results demonstrate the performance gain from the \ac{EWPT} structure (\sys{}-Static vs. Fabric-IoT) and the additional gain from dynamic optimization (\sys{} vs. \sys{}-Static).}
	\label{fig:perf_overview}
\end{figure*}

This paper evaluated the three schemes—Fabric-IoT, \sys{}-Static, and \sys{}—across all 15 test cases, measuring system throughput (Transactions Per Second, TPS) and average authorization latency. The results for the seven factor groups are summarized in Fig.~\ref{fig:perf_overview}.

The analysis of results provides clear answers to the \ac{RQ}:
\begin{enumerate}[label=(\textbf{\roman*}), wide]
    \item \textbf{Answer to RQ2 (Performance Advantage \& Source):} \sys{} achieves substantial performance improvements.
    \begin{itemize}
        \item \textbf{\ac{EWPT} Contribution:} \sys{}-Static consistently and significantly outperforms Fabric-IoT (e.g., $\sim$10$\times$ higher throughput, $\sim$90\% lower latency). This confirms that the EWPT structure itself, by reducing authorization to an $O(m)$ path traversal, is the primary source of performance gain.
        \item \textbf{Dynamic Optimization Contribution:} \sys{} further enhances performance beyond \sys{}-Static (e.g., $\sim$11$\times$ throughput gain over Fabric-IoT). This demonstrates the synergistic effect of the dynamic optimization module, which continuously adapts the \ac{EWPT} to current access patterns, ensuring optimal performance.
    \end{itemize}
	\item \textbf{Answer to RQ3 (Scalability and Resilience):} \sys{} demonstrates excellent scalability and resilience.
    \begin{itemize}
        \item \textbf{Scalability:} Under increasing system scale (Fig.~\ref{fig:perf_overview}a,b,c), \sys{} maintains stable or even improving performance, while the baselines show degradation. This advantage stems from its ability to dynamically utilize historical access information to continuously optimize the authorization path structure within the \ac{EWPT}, thereby enhancing decision-making efficiency. The system's feedback loop becomes more effective at identifying optimal paths as more data becomes available.
        \item \textbf{Resilience to Policy Growth:} With increasing policy complexity (Fig.~\ref{fig:perf_overview}d), \sys{} shows superior robustness (only $\sim$10\% drop vs. $\sim$40\% for Fabric-IoT).
        \item \textbf{Resilience to Attribute Explosion:} Against increasing attribute dimensions (Fig.~\ref{fig:perf_overview}e,f,g), \sys{} exhibits the smallest performance decline ($<$13\% throughput drop). The dynamic prioritization in the \ac{EWPT} effectively mitigates the performance impact of attribute explosion.
    \end{itemize}
\end{enumerate}
\ZJ{
While the current experimental scale comprehensively covers typical IoT deployment scenarios \cite{Ahmed2024DatasetFA}, the scalability of \sys{} under even higher loads warrants discussion. The \ac{EWPT} structure fundamentally reduces the policy matching complexity to $O(m)$, where $m$ is the number of attributes in a request, making it independent of the total number of policies. This theoretical guarantee, combined with the observed trends in Fig.~\ref{fig:perf_overview}a,b,c where performance remains stable or improves with increasing numbers of subjects, objects, and requests, indicates a strong inherent scalability, suggesting that \sys{} is well-positioned to handle larger-scale deployments.
}

\textbf{Summary for RQ2 \& RQ3:} \sys{} achieves significant performance gains over traditional \ac{ABAC}, primarily due to the \ac{EWPT} structure, with a further boost from dynamic optimization. It is highly scalable and resilient, maintaining efficient performance even as system size and complexity grow, making it suitable for large-scale, dynamic environments. \ZJ{Actually, the synergy between anonymity and efficiency in \sys{} stems from a fundamental insight: highly anonymous attributes (those with widespread occurrence in the subject space) naturally form efficient matching paths in the \ac{EWPT}. By prioritizing these attributes, \sys{} simultaneously enhances both privacy protection and matching efficiency, creating a virtuous cycle rather than a trade-off. Combined with regular weight updates and \ac{EWPT} reconstruction, when policies are modified, the resulting changes in access patterns are captured in subsequent optimization cycles, automatically adapting the authorization structure, ensures continuous optimization under evolving policy environments.
}

%% file: 09.Conclusion.tex
\section{Discussion}
\label{sec:discussion}

\subsection{Limitations}
\subsubsection{Limitations of the Anonymity Metric}
The $\mathcal{E}_{req}$ metric's foundational assumption of uniform adversarial priors proves vulnerable to real-world attackers with auxiliary knowledge, including non-uniform prior distributions, exploitable attribute correlations within credentials, and behavioral/temporal side-channels in request patterns. Although increasing the anonymity threshold can alleviate this risk, this approach imposes significant throughput costs, highlighting a fundamental privacy-performance tradeoff.

\subsubsection{Threat Model Implications}
Relaxing blockchain trust assumptions reveals critical vulnerabilities: compromised nodes accessing sensitive state (\ac{AAH} Pool, attribute matrix $M$), malicious administrators bypassing anonymity safeguards despite signature protections, and timing-based de-anonymization through transaction metadata. These threats define precise security boundaries for \sys{}: while the framework provides robust protection against external passive adversaries, addressing advanced internal threats requires orthogonal mitigation strategies with substantial performance tradeoffs beyond its current scope.

\ZJ{
\subsection{Practical Deployment Considerations}
\label{subsec:deployment}
This section further discusses the practical usability and integration of \sys{} in real-world deployments. The modular chaincodes \ac{AQC}/\ac{AIC}/\ac{PDC} in \sys{} are developed in Go, successfully deployed on the widely adopted Hyperledger Fabric platform, demonstrating excellent compatibility. The deployment overhead primarily consists of initial attribute space configuration and access control policy downloads during operation, without incurring additional costs. For \ac{ABAC} systems implemented on Fabric (e.g., \cite{Liu2020FabriciotAB}), integrating \sys{} mainly involves replacing existing authorization logic with our contracts, exhibiting backward compatibility. Crucially, the automated execution feature of chaincodes introduces no supplementary overhead. Consequently, the proposed \sys{} possesses genuine deployability.
}

\subsection{Future Work}
\label{sec:future_work}

In \textbf{Enhanced Anonymity Metrics}, future research will develop Bayesian anonymity metrics for privacy quantification under adversarial knowledge, alongside integrating differential privacy with calibrated noise injection into \ac{AAH} pools and $\mathcal{SS}_c$ spaces to strengthen guarantees against auxiliary information. In \textbf{Trusted System Implementation}, \ZJ{trusted execution environments will be explored to secure critical computations and \ac{ZKP} will be investigated} for enabling credential-based authorization without identity exposure, thereby eliminating the need for on-chain anonymity calculations. In \textbf{Cross-Domain Adaptation}, research directions include designing federated anonymity estimation for multi-domain scenarios without data sharing, and implementing online learning mechanisms to dynamically adapt to concept drift and evolving adversarial tactics.

\section{Conclusion}
\label{sec:conclusion}

This paper has presented \sys{}, a novel framework that effectively tackles the dual challenges of privacy preservation and authorization efficiency in blockchain-based attribute-based access control. By introducing a quantifiable $(r, t)$-anonymity model for continuous privacy assessment and an \ac{EWPT} for privacy-aware policy optimization, \sys{} achieves a breakthrough balance between these traditionally conflicting goals. Extensive experimental results demonstrate that this framework not only maintains strong anonymity guarantees but also significantly enhances performance, yielding up to 11× higher throughput and 87\% lower latency compared to state-of-the-art alternatives, thereby enabling practical and secure fine-grained access control for decentralized applications.

%% file: Appendix.tex
\appendix

\section{Full Security Proof of Theorem~\ref{thm:main-security}}
\label{app:security-proof}

This appendix provides the complete formal proof for Theorem~\ref{thm:main-security}. \ZJ{The focus is placed} on the \textit{Request Anonymity} game $\mathsf{Exp}_{\mathcal{A},\Pi}^{\mathsf{Req-Anon}}(1^\lambda)$. The proof for \textit{Unlinkability} follows a similar structure.

\subsection{Proof Setup}

Let $\mathcal{A}$ be a \ac{PPT} adversary with non-negligible advantage $\epsilon(\lambda) = \mathsf{Adv}_{\mathcal{A},\Pi}^{\mathsf{Anon}}(\lambda)$ in the request anonymity game. \ZJ{Two simulators, $\mathcal{S}_1$ and $\mathcal{S}_2$, are constructed} such that one of them can break the underlying security assumption.

The \ac{EUF-CMA} challenger for signature scheme $\Sigma$ provides $\mathcal{S}_1$ with a public key $pk^*$ and grants $\mathcal{S}_1$ access to a signing oracle $\mathsf{Sign}(sk^*, \cdot)$ for the corresponding secret key $sk^*$.
The \ac{DL} challenger provides $\mathcal{S}_2$ with a group description $(\mathbb{G}, p, g)$ and a challenge element $h = g^a \in \mathbb{G}$.

Both simulators $\mathcal{S}_1$ and $\mathcal{S}_2$ must simulate the environment for $\mathcal{A}$, including the system setup, the subject space $S$, the request history $R$, and answering all of $\mathcal{A}$'s queries.

\subsection{Simulator Construction: $\mathcal{S}_1$ (\ac{EUF-CMA} Reduction)}

Simulator $\mathcal{S}_1$ interacts with the \ac{EUF-CMA} challenger and adversary $\mathcal{A}$ as follows:

\begin{enumerate}
	\item \textbf{Initialization:} $\mathcal{S}_1$ receives $pk^*$ from the challenger. It generates the system parameters, the subject space $S$, and the attribute space $A$. It assigns the public key $pk^*$ to one of the subjects, say $s^*$, chosen uniformly at random. For all other subjects $s_i \neq s^*$, $\mathcal{S}_1$ generates key pairs $(pk_i, sk_i)$ honestly.
	\item \textbf{Oracle Simulation:}
	\begin{itemize}
		\item \textbf{Signing Oracle:} When $\mathcal{A}$ requests a signature for subject $s_i$ on a credential $c$:
		\begin{itemize}
			\item If $s_i \neq s^*$, $\mathcal{S}_1$ uses the known $sk_i$ to compute $\sigma \gets \mathsf{Sign}(sk_i, c)$.
			\item If $s_i = s^*$, $\mathcal{S}_1$ queries the \ac{EUF-CMA} challenger's signing oracle on the message $c$ and relays the signature $\sigma^*$ back to $\mathcal{A}$.
		\end{itemize}
		\item \textbf{Request History $R$:} $\mathcal{S}_1$ can perfectly simulate $R$ by generating requests for any subject using the signing oracle as described above.
	\end{itemize}
	\item \textbf{Anonymity Challenge:} When $\mathcal{A}$ outputs the challenge subjects $(s_0, s_1)$ and the target credential $c^*$:
	\begin{itemize}
		\item If $s^* \notin \{s_0, s_1\}$, $\mathcal{S}_1$ aborts.
		\item Otherwise, assume w.l.o.g. $s^* = s_0$. $\mathcal{S}_1$ queries the \ac{EUF-CMA} challenger for a signature $\sigma^*$ on $c^*$. It then constructs the challenge request $req_b = (\sigma^*, c^*, o, op, env)$ and sends it to $\mathcal{A}$.
	\end{itemize}
	\item \textbf{Output:} If $\mathcal{A}$ wins the game (outputs $b' = b$), and the credential $c^*$ was never queried to the signing oracle for subject $s_0$, then the pair $(c^*, \sigma^*)$ is a valid forgery. $\mathcal{S}_1$ outputs it to win the \ac{EUF-CMA} game.
\end{enumerate}

\noindent
\textbf{Probability Analysis for $\mathcal{S}_1$:}
The probability that $\mathcal{S}_1$ does not abort is at least $1/|S|$, which is non-negligible. If $\mathcal{A}$ wins with advantage $\epsilon(\lambda)$, and the abort event is independent, then:
\[
\mathsf{Adv}_{\Sigma}^{\mathsf{EUF-CMA}}(\mathcal{S}_1) \geq \frac{\epsilon(\lambda)}{|S|} - \mathsf{negl}(\lambda)
\]
where the negligible term accounts for the probability of guessing the signature without the key.

\subsection{Simulator Construction: $\mathcal{S}_2$ (\ac{DL} Reduction)}\label{app-signature}

Simulator $\mathcal{S}_2$ interacts with the \ac{DL} challenger and adversary $\mathcal{A}$.

\begin{enumerate}
	\item \textbf{Initialization:} $\mathcal{S}_2$ receives $(\mathbb{G}, p, g, h=g^a)$ from the challenger. It sets up the system. For the challenge subject $s^*$, it sets $pk^* = h$ (embedding the \ac{DL} challenge as the public key). For all other subjects, it generates keys honestly $(pk_i, sk_i) = (g^{sk_i}, sk_i)$.
	\item \textbf{Oracle Simulation:} $\mathcal{S}_2$ must simulate signatures for $s^*$ \textit{without knowing} $a = \mathsf{dlog}_g(h)$.
    \begin{itemize}
        \item For a \textit{deterministic} signature scheme, the signature function $\sigma = \mathsf{Sign}(sk, m)$ is a deterministic function of the private key $sk$ and the message $m$. Consequently, for a given credential $c$ and public key $pk^* = h = g^a$, the valid signature $\sigma_s(c)$ is uniquely determined. Simulating a valid signature $\sigma$ for $s^*$ without knowledge of $a$ is therefore equivalent to computing the function $\mathsf{Sign}(a, c)$ without the input $a$, which violates the fundamental security requirement of the signature scheme. This presents a fundamental barrier for reduction in the standard model.
        \item This barrier can be circumvented in the \textit{\ac{ROM}}. Here, $\mathcal{S}_2$ can simulate signatures for $s^*$ by programming the random oracle. For a credential $c$, it can generate a random signature $\sigma$ and program the random oracle such that $\mathsf{Verify}(pk^*=h, c, \sigma) = \mathsf{True}$. This simulation technique is well-established for schemes like ECDSA~\cite{wong2023real} within the \ac{ROM}.
    \end{itemize}
    $\mathcal{S}_2$ simulates the rest of the environment perfectly.
	\item \textbf{Anonymity Challenge:} $\mathcal{A}$ outputs $(s_0, s_1)$. If $s^* \notin \{s_0, s_1\}$, abort. Otherwise, $\mathcal{S}_2$ generates the challenge request for $s^*$ using the signature simulation technique above.
	\item \textbf{Extraction:} If $\mathcal{A}$ wins the game, it has demonstrated an ability to link the signature to the specific public key $h$. $\mathcal{S}_2$ can analyze the transcripts and the random oracle queries made by $\mathcal{A}$ to extract the discrete logarithm $a$. The exact extraction algorithm depends on the specific signature scheme used.
\end{enumerate}

\noindent
\textbf{Probability Analysis for $\mathcal{S}_2$:}
The success probability of $\mathcal{S}_2$ depends on the extractability of the signature scheme in the \ac{ROM}. For a well-designed scheme, if $\mathcal{A}$ has advantage $\epsilon(\lambda)$, then:
\[
\mathsf{Adv}_{\mathbb{G}}^{\mathsf{DL}}(\mathcal{S}_2) \geq \frac{\epsilon(\lambda)}{\poly(\lambda)} - \mathsf{negl}(\lambda)
\]
where $\poly(\lambda)$ is a polynomial factor representing the cost of extraction (e.g., related to the number of random oracle queries).

\subsection{Conclusion of the Proof}

Combining the two cases, the advantage of any \ac{PPT} adversary $\mathcal{A}$ against the anonymity of $\Pi$ is bounded by:
\[
\mathsf{Adv}_{\mathcal{A},\Pi}^{\mathsf{Anon}}(\lambda) \leq |S| \cdot \mathsf{Adv}_{\Sigma}^{\mathsf{EUF-CMA}}(\lambda) + \poly(\lambda) \cdot \mathsf{Adv}_{\mathbb{G}}^{\mathsf{DL}}(\lambda) + \mathsf{negl}(\lambda)
\]
Since both $\mathsf{Adv}_{\Sigma}^{\mathsf{EUF-CMA}}(\lambda)$ and $\mathsf{Adv}_{\mathbb{G}}^{\mathsf{DL}}(\lambda)$ are negligible by assumption, and $|S|$ and $\poly(\lambda)$ are polynomials, the entire expression is negligible. This completes the proof of Theorem~\ref{thm:main-security}.